\documentclass[aps,pra,reprint,floatfix,superscriptaddress,longbibliography,nofootinbib]{revtex4-2}
\usepackage{tabularx} 
\usepackage{amsmath}  
\usepackage{amssymb}
\usepackage{dsfont}
\usepackage{blindtext}
\usepackage{appendix}
\usepackage{xfrac}
\usepackage{graphicx} 
\usepackage[final]{hyperref} 
\usepackage{braket}
\usepackage{multirow}
\usepackage[linesnumbered,ruled,vlined]{algorithm2e}
\usepackage{tikz}
\usepackage{svg}
\usepackage{subcaption}
\usetikzlibrary{quantikz2, arrows, positioning, shapes.geometric}
\hypersetup{
	colorlinks=true,       
	linkcolor=blue,        
	citecolor=blue,        
	filecolor=magenta,     
	urlcolor=blue         
}
\DeclareMathOperator*{\argmin}{arg\,min}

\begin{document}

\title{Unitary Dilation Strategy Towards Efficient and \\ Exact Simulation of Non-Unitary Quantum Evolutions}
\author{Aman Mehta}
\affiliation{Department of Electrical and Computer Engineering, University of California, Los Angeles, California, USA}
\author{Scott E. Smart}
\affiliation{College of Letters and Science, University of California, Los Angeles, California 90095}

\author{Joseph Peetz}
\affiliation{Department of Physics and Astronomy, University of California, Los Angeles, California 90095}

\author{David A. Mazziotti}
\affiliation{Department of Chemistry and The James Franck Institute, The University of Chicago, Chicago, IL 60637}%

\author{Prineha Narang}
\affiliation{Department of Electrical and Computer Engineering, University of California, Los Angeles, California, USA}  
\affiliation{College of Letters and Science, University of California, Los Angeles, California 90095}

\date{\today}

\begin{abstract}
	Simulating quantum systems with their environments often requires non-unitary operations, and mapping these to quantum devices often involves expensive dilations or prohibitive measurement costs to achieve desired precisions. 
    Building on prior work with a finite-differences strategy, we introduce an efficient and exact single-ancilla unitary decomposition technique that addresses these challenges. Our approach is based on Lagrange-Sylvester interpolation, akin to analytical differentiation techniques for functional interpolation. 
    As a result, we can \emph{exactly} express any arbitrary non-unitary operator with no finite approximation error using an easily computable decomposition. 
    This can lead to several orders of magnitude reduction in the measurement cost, which is highly desirable for practical quantum computations of open systems.  
\end{abstract}

{
\maketitle
}
 

 \section*{Introduction}

Open quantum systems arise when a real quantum system interacts with its environment and the external degrees of freedom have a non-negligible effect on the dynamics of the system. Such systems are encountered ubiquitously across physics, and their simulation has potential applications in, but not limited to, condensed matter \cite{intro_oqs_cmp1, intro_oqs_cmp2, intro_oqs_cmp3, intro_oqs_cmp4_thermo}, quantum chemistry \cite{intro_oqs_chem1}, thermodynamics \cite{intro_oqs_cmp4_thermo, intro_oqs_thermo2}, metrology \cite{intro_oqs_metro1, intro_oqs_metro2} and quantum information \cite{intro_oqs_qis1, intro_oqs_qis2, intro_oqs_qis3, intro_oqs_qis4, intro_oqs_qis5}. Though operations on quantum computers are strictly unitary, they are a promising candidate to carry out open dynamics simulations through non-unitary evolution, and several proposed quantum algorithms may provide computational advantages in directly calculating the evolution of density matrices \cite{theory_oqs_book, lindblad_master_eq, cpds_oqs}. A common procedure to realize these using quantum computation is through dilation \cite{Stinespring_dilation, dilation_paulsen, dilation_sz_nagy}, i.e. encoding a unitarized expansion within a larger Hilbert space. Notable strategies include simulating the Lindbladian \cite{lindbladian_trotter1, lindbladian_trotter2, lindbladian_trotter3, lindbladian_trotter4, lindbladian_sim1, lindbladian_sim2, lindbladian_efficient_simulation},  Sz.-Nagy dilation \cite{intro_sznagy_1, intro_sznagy_2, intro_sznagy_3, sz-nagy_nmr, sz-nagy_fmo_complex, sz-nagy_david-pri}, singular value decomposition \cite{intro_svd_1, intro_svd_2} and other unitary decompositions \cite{intro_lcu_1, original_paper, two_unitary_decomposition}.
Previous work by Schlimgen \emph{et al.} \cite{original_paper} introduced a straightforward unitary decomposition for an arbitrary non-unitary operator using a first-order expansion. This method, while being straightforward to implement, can have a prohibitively high measurement cost. 

Here, we introduce an interpolation-based strategy that is similarly exact and has a fundamentally different measurement scaling, resulting in several orders of magnitude savings in resources. In principle, the distinction between the two methods is similar to the difference between numerical and automatic differentiation, which is demonstrated in the parameter shift rule \cite{theory_crooks_psr, general_psr}. Beyond the minimal case, we can apply this to arbitrary dimensional non-unitary operators using dilation strategies, including recently introduced stochastic approaches \cite{peetz_scu, chakrabortyImplementingAnyLinear2023}. The overhead is lower bounded not by the dimension but instead by the maximal eigenvalue of the decomposition, which for quantum channels is upper bounded by one. We show that in most cases this lower bound is practically attainable and as a result, the method is highly scalable. Later, we discuss the applications of this approach for simulating open quantum systems.

\section*{Theory} 
The dynamics of quantum operations can be described using Completely Positive Trace Preserving (CPTP) Maps \cite{Nielsen_Chuang_2010}. These can be expressed in a Kraus operator sum formalism, where a set of $K$ operators $M_k$, evolve a density matrix $\rho$ as $\tilde{\rho} = \sum_{k=1}^K M_k\rho_0 M_k^\dag$. These possess a trace normalization condition, $\sum_{k=1}^K M_k^\dagger M_k = I$, and are not necessarily unitary, requiring us to encode non-unitary operators with unitary quantum operations. 

\subsection*{Unitary Decomposition of Quantum Operators} 

Any operator $M$ can be expressed as a sum of Hermitian and anti-Hermitian components, $S$ and $A$, such that:
\begin{align}
    S,A &= \frac{1}{2}{(M + M^\dag)}, \frac{1}{2}{(M - M^\dag)}.
    \label{eq:SA_decomposition}
\end{align}
These matrices can be treated as the generators of unitary operators and thus can be written as the sum of these unitaries, in the first-order expansion:
\begin{align}
    S &=  \lim_{\epsilon \to 0} \frac{\iota}{2\epsilon}(e^{-\iota\epsilon S} - e^{\iota\epsilon S})  \\
    A &= \lim_{\epsilon \to 0} \frac{1}{2\epsilon}(e^{\epsilon A} - e^{-\epsilon A})
    \label{eq:A_expansion_approximate}
\end{align}
where $\epsilon$ is the expansion parameter and $S_m, S_p, A_m, A_p$ are unitary matrices that are encoded in a block diagonal form in the operator $U$ to prepare the final state.

The power of this method lies in its simplicity, as it allows us to decompose any arbitrary operator into a sum of at most four unitaries which can be implemented easily using a linear combination of unitaries \cite{LCU_Weibe}. The final result is then classically rescaled by factors of $\epsilon$, generally less than 1. Thus, a quantum channel can simulated by continuously applying the decomposition on all Kraus operators at each time step. The largest errors in the first order expansion are of the order $O(\epsilon^2)$. However, this leads to a substantial increase in measurement costs as the observables must also be rescaled. Naively,  the variance is proportional to $O(1/\epsilon^2)$, which requires further measurements or extrapolation. 



\subsection*{Exact Unitary Expansion of Arbitrary Matrices}
\label{subsec:exact_unitary_expansion}
While techniques like the parameter-shift rule \cite{lagrange_interpolation_psr} can be applied for simple generators in terms of function evaluations, generalizing these to more complex eigenvalue expressions or matrix functions is more challenging. Here we introduce a procedure for finding analytical expressions of matrix functions using matrix interpolation. 

Within the scheme of Sylvester-Lagrange interpolation \cite{19_dubious_ways_lagrange, Tarantola_2006, lagrange_degen1}, we can express any analytic function, $f$, of an $N-$dimensional square matrix $M$ as a finite matrix polynomial 
\begin{equation}
    f(M) = \sum_{j=0}^{N}g_j(\mathbf{\lambda})M^j,
\end{equation} 
where $\mathbf{\lambda}$ are the eigenvalues of $M$ and $g_j$ are the Frobenius covariants calculated using the Cayley-Hamilton theorem. Letting $f(M) = e^{-i\mu M}$, and evaluating this function at $N$ different points $\mu$, we can assemble a matrix of coefficients and solve for the first order term yielding the target interpolations:
\begin{align}
    S &= \sum_i^N c^{(s)}_{i} e^{-\iota\mu^{(s)}_{i}S} \label{eq:expansionS} \\
    A &= \sum_i^N c^{(a)}_{i} e^{-\mu^{(a)}_{i} A} 
    \label{eq:expansionA}
\end{align}
where $\vec{c} {}^{(s/a)} = \{c^{(s/a)}_{i}\}\in \mathcal{C}^d$ is the solution of the linear system of equations. Since the unitaries and their generators are simultaneously diagonalizable, we can also write Eqs. \eqref{eq:expansionS}-\eqref{eq:expansionA} in the eigenvalue basis as 
\begin{align}
    \lambda_i &= \sum_j^n c_je^{-\iota\mu_j\lambda_i} \qquad \forall i \in [0,1,..., n]
    \label{eq:lambdai=sumj_cj_exp-iota_muj_lambdai}
\end{align}
for $n\leq N$ unique eigenvalues and similarly solve the linear systems. The latter approach does not require the calculation of the Frobenius covariants and operationally is simpler, although both assume we have access to the eigenvalue decomposition. As a result, we can decompose any non-unitary operator into a sum of at most $2N$ unitaries.

Despite its exactness, to implement this on a quantum computer, we use an encoding that scales with the $l_1$ norm of $\vec{c}$ \cite{LCU_Weibe,peetz_scu}. We can establish a lower bound on the coefficient vector norm by applying the triangle inequality on the right-hand side of Eq. (\ref{eq:lambdai=sumj_cj_exp-iota_muj_lambdai}) to get
\newline
\begin{align}
     \sum_j^n |c_j e^{-\iota \mu_j\lambda_i}| &\geq |\sum_j^n c_j e^{-\iota \mu_j\lambda_i}| \qquad \forall i \nonumber \\
      \implies \qquad  |\vec{c}\,|_1 &\geq |\vec{\lambda}|_{\infty}  \label{eq:inequal}
\end{align}
where $|\vec{\lambda}|_{\infty}$ is the maximum unsigned eigenvalue of the matrix. An ideal interpolation minimizes the norm of the interpolation coefficients over a span of $n$ points:
\begin{equation}
    \vec{\mu}^* = \argmin_{\vec{\mu}} | \mathbf{E}(\vec{\mu})^{-1} \vec{\mathbf{\lambda}}|_1
    \label{eq:EC=Lambda},
\end{equation}
where $\mathbf{e}_{i,j} = e^{-\iota\mu_j\lambda_i}$ are the elements of a matrix $\mathbf{E}$ and $ \vec{\mathbf{\lambda}}$ is a vector of eigenvalues.

\begin{figure}[t!]
    \includegraphics[width=\columnwidth]{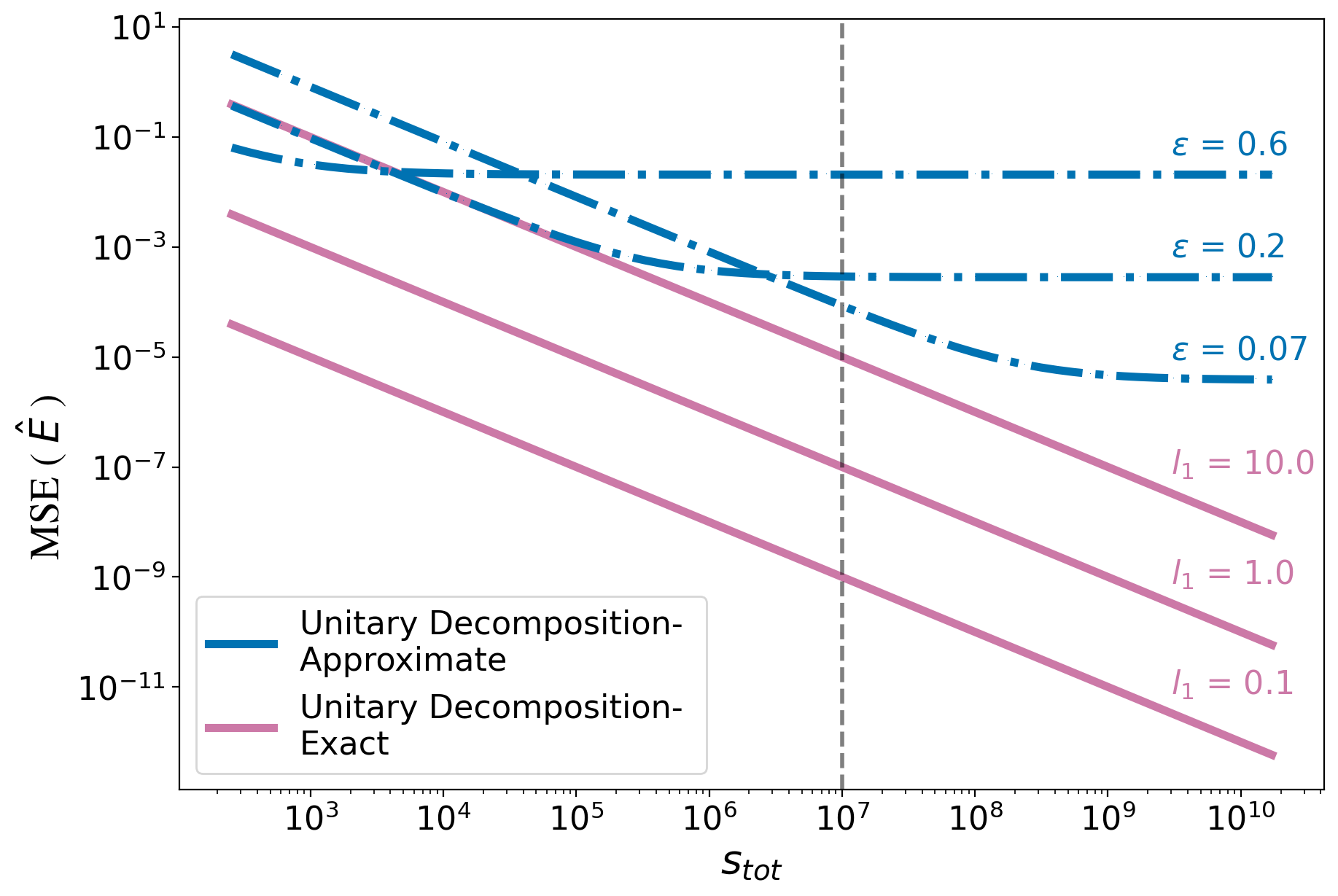}
    \caption{Mean square error (MSE) versus total shots ($s_{\text{tot}}$) for the approximate (blue dashed) and exact (pink solid) unitary decompositions for several $\epsilon$ and $l_1$ norms, respectively. The flat regions correspond to non-zero bias in the first-order truncations, which is not present in the exact approach.}
    \label{fig:mse_vs_stot}
\end{figure}

As an example, for $n=2$, constraining $\mu_1 = \mu_2$ we can show that:
\begin{equation}
    \mu^* = \pm\frac{2}{\lambda_0 - \lambda_1}\arctan{\sqrt{\bigl|\dfrac{\lambda_0 - \lambda_1}{\lambda_0 + \lambda_1}\bigr|}}
\end{equation}
which yields a minimum $l_1$ norm of 
\begin{equation}
    |\vec{c}|_{1}^* = \frac{1}{2}(|\lambda_0 - \lambda_1| + |\lambda_0 + \lambda_1|) = \max{(|\lambda_0|, |\lambda_1|)}
\end{equation}
saturating the previous inequality. The entire proof is detailed in Appendix \ref{sec:interpolation}. For larger dimensions, it is challenging to analytically prove that we can reach this lower bound, and ultimately it depends on the norm of the inverse matrix, which itself depends on $\vec{\mu}$. Practically, we can use techniques from numerical optimization to solve the above minimization problem.

\begin{figure}[t]
    \includegraphics[width=0.95\columnwidth]{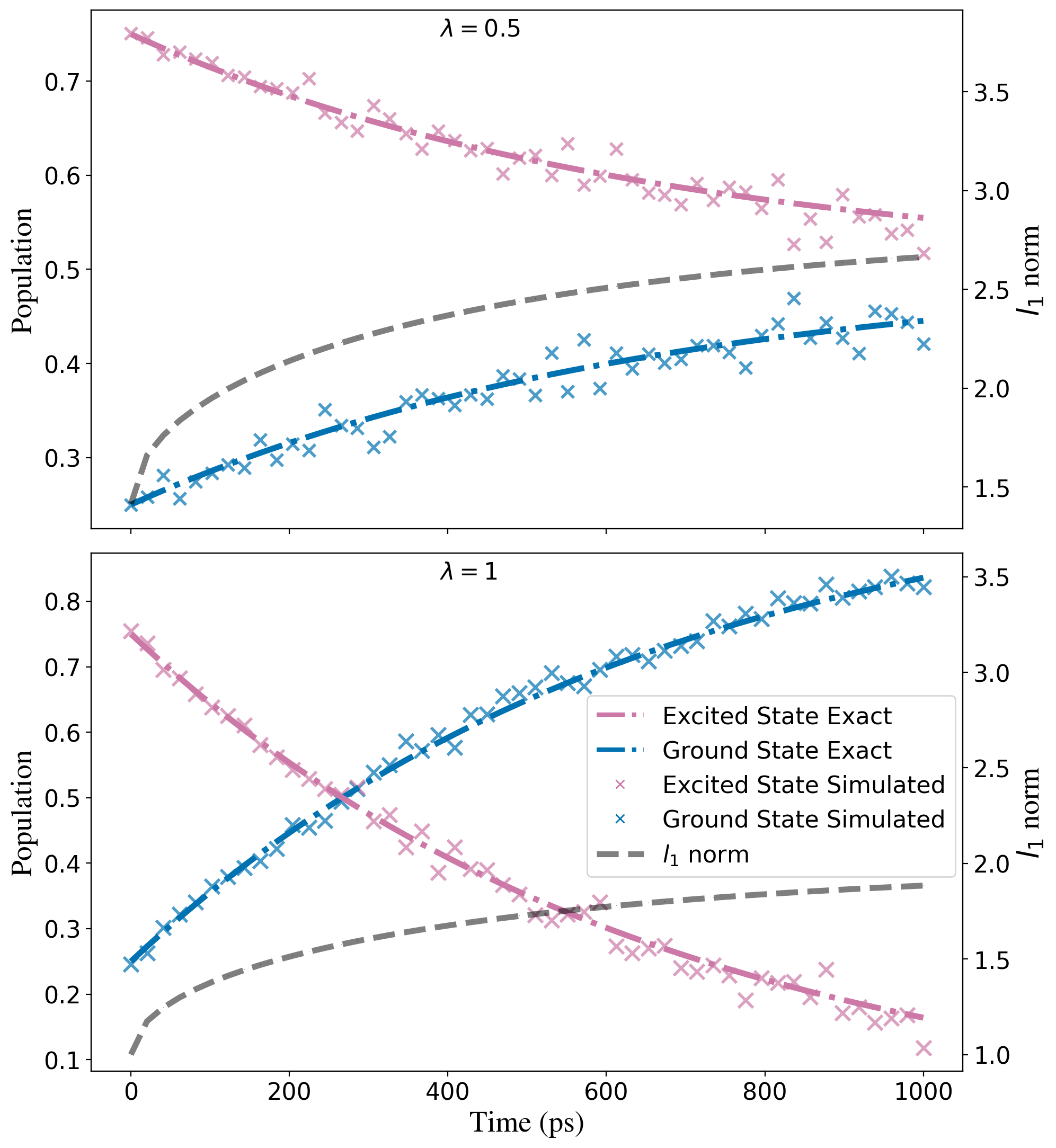}
        
    \caption{Excited state population (pink) decaying to ground state (blue) for room temperature (top), $\lambda=0.5$, and near zero temperature (bottom), $\lambda=1$. The dotted lines represent the exact evolution and $\times$ are obtained with $2^{11}$ shots of a noiseless simulator, around a 1000-fold decrease in shots from the approximate decomposition. The black curve is the $l_1$ norm of the computed coefficients for the complete Kraus expansion at a given time. }
    \label{fig:adc_results}
    
\end{figure}

Importantly, here there is no dependence on $\epsilon$, and in Fig. \ref{fig:mse_vs_stot} we demonstrate the mean squared error (MSE) of estimating a generic observable for the approximate and exact methods. While Richardson extrapolation \cite{Richardson_Gaunt_1927, richardson_error} can be used to reduce bias with only a few points, each point must be extracted with a large number of samples.


The overall approach is as follows. For each Kraus operator we calculate the spectra of $S$ and $A$ and solve the optimization problem in Eq.~\eqref{eq:EC=Lambda}. To implement the circuits, we can use a standard dilation, such as linear combinations of unitaries (LCU) \cite{LCU_Weibe,two_unitary_decomposition}, or more resource-efficient approaches such as the stochastic combination of unitaries (SCU) or single-qubit LCU \cite{peetz_scu, chakrabortyImplementingAnyLinear2023}. After performing the dilation, we calculate the target observable through the appropriate state reconstruction formalism. The pseudocode of the algorithm for the SCU approach can be found in Algorithm \ref{alg:unitary_decomposition} in Appendix \ref{sec:method}.

The LCU requires $\log_2 (n) + 1$ ancilla qubits circuits and controlled operations for each Kraus operator but also can take advantage of super-normalized dilations. Using SCU however, we require only a single qubit with a variance proportional to $L^2$, where $L = \sum_{k=1}^K |\vec{c}~^k|_1$. This is in contrast to the LCU variance of the approximate decomposition of $\frac{K}{\epsilon^2}$. We provide a more detailed comparison of these dilation and unitary decomposition strategies in Appendix \ref{sec:dilations}. 

\section*{Results}

We first demonstrate our approach using the dynamics of a two-level amplitude damping channel \cite{generalized_adc, original_paper} at zero and finite temperature. The amplitude damping channel is a non-unitary operation that models the physical processes such as spontaneous emission, by which the population in the excited state decays to the ground state.
 
We used the initial state $\rho_0= \frac{1}{4}I + \frac{1}{2}|1\rangle \langle 1 |$ with a standard amplitude damping channel (see \cite{adc_krausmap,original_paper, huQuantumAlgorithmEvolving2020a}) and set the decay rate $\gamma = 1.52 \times 10^{9} \text{s}^{-1}$. The specific Kraus operators are $M_0 = \sqrt{\lambda}(|0\rangle \langle 0|  + \sqrt{e^{-\gamma t}} \ket{1}\bra{1})$, $M_1 = \sqrt{\lambda}\sqrt{1-e^{-\gamma t}} \ket{0}\bra{1} $, $M_2 =  \sqrt{1-\lambda}(\sqrt{e^{-\gamma t}}\ket{0}\bra{0} + \ket{1}\bra{1}) $, $M_3 = \sqrt{1-\lambda}\sqrt{1-e^{-\gamma t}} \ket{1}\bra{0}$, where $\lambda = 1/(1+ e^{-1/k_BT})$ accounts for the distribution shift due to temperature.

We calculate the populations for different temperature values sampling from a noiseless quantum simulator. The results can be seen in Fig. \ref{fig:adc_results}. The dilations are carried out using the SCU method, detailed in Appendix \ref{sec:method}, and result in an increased measurement overhead. The simulation results marked by $\times$ follow the exact result while maintaining a total measurement overhead which is less than $\sqrt{K}$ for all times. 
        
\begin{figure}[t]
    \includegraphics[width=\columnwidth]{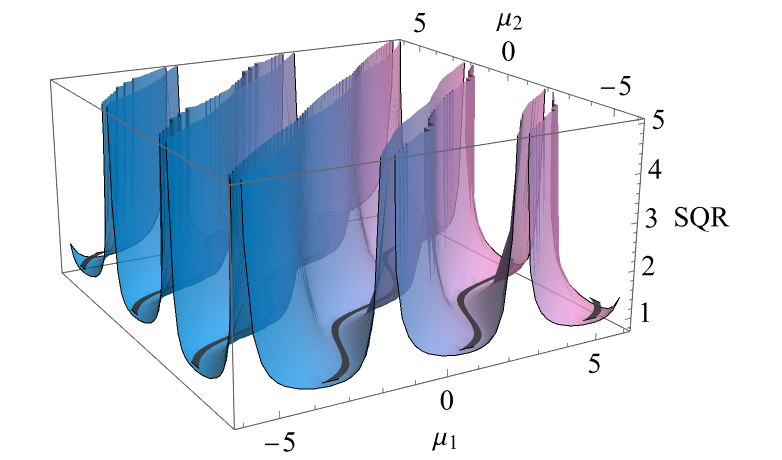}
    \caption{Interpolation cost in terms of the $l_1$ norm for  parameters $\mu_1$ and $\mu_2$ in a 2-dimensional system. Despite the landscape being highly non-convex, we can rapidly find solutions using techniques from classical optimization.}
    \label{fig:opt_landscape}
\end{figure}

To investigate the performance of our approach for larger non-trivial systems, we first introduce a new metric, the Solution Quality Ratio (SQR), defined as the ratio of coefficient $l_1$ norm to the maximum eigenvalue of $M$, and which is ideally 1. Plotting the SQR against the input parameters in Fig. \ref{fig:opt_landscape},  for a 2-dimensional system, provides an insight into the topology of the optimization landscape and the solution space.

Consequently, in our analysis, we observe that the performance is sensitive to the optimization strategy employed. Substantial variables that affect the SQR include the initial parameters selection, choice of optimizer, and dimensionality of the problem. Running the algorithm for multiple randomly generated sets of Kraus Maps, we find that the Sequential Least Squares Programming (SLSQP) method generally outperforms other optimization methods. For arbitrary $n$ the classical optimization has iterations scaling as $n^3$, which is exponential in the dimension of the Kraus maps.

\begin{figure}[t]
    \includegraphics[width=\columnwidth]{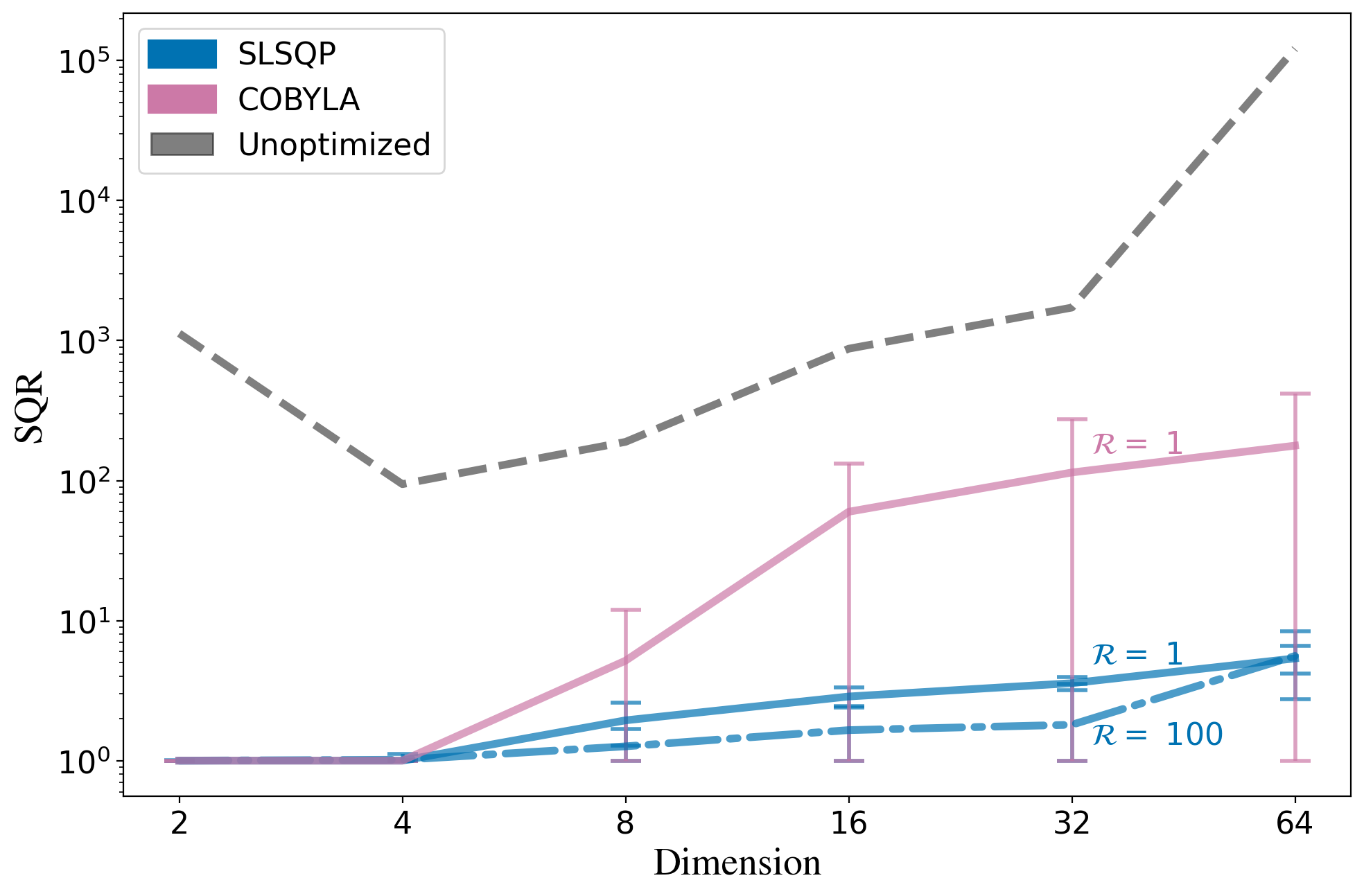}
    \caption{Demonstration of optimization strategy on the measurement cost for randomly sampled eigenvalue distributions $\mathbb{R}^d \in [0,1]$ for varying dimensions. The performance of the problem depends closely on the optimizer (here showing SLSQP, COBYLA, and randomly selected points) and initialization strategy (scaled by $\mathcal{R}$), but exhibits very reasonable scaling for moderate dimensions.}
    \label{fig:max_eigenvalue_l1_ratio}
\end{figure}

Based on these observations, we investigated a heuristic subroutine that performs a low-cost shallow optimization over differing initial parameter scale factors (denoted by $\mathcal{R}$) and optimizers and then performs an optimization over the optimal scaling factor. As a result, even for relatively large dimensions for random eigenvalue distributions, we obtain very reasonable measurement overhead requirements. A highlight of these results are shown in Fig.~\ref{fig:max_eigenvalue_l1_ratio}, where we show the variability of the SQR for exponentially increasing dimension size with respect to optimization strategies.

\vspace{0.4cm}
\section*{Conclusions and Outlook}

The present work provides a straightforward way to represent any non-unitary operator as an exact linear combination of unitary operators with no dependence on a finite difference factor or on the number of Kraus terms. Using a stochastic approach, this dilation can be accomplished with a single ancilla qubit. The particular decomposition we find is not unique though solutions can be obtained via optimization of an interpolation problem. The resulting estimator is unbiased and offers a way to simulate non-unitary processes exactly, removing the finite difference factors of previous work.

While we demonstrated the scalability of the interpolation for larger Kraus maps, Kraus operators for physical systems are commonly (though not exclusively) local with few unique eigenvalues, though may be applied to arbitrarily large density matrices. For much larger Kraus maps, improved interpolation schemes or more resource-intensive quantum algorithms may be needed. Regardless, the algorithm is promising for near-term quantum devices and can easily be applied to problems in physics and chemistry involving open systems, which is appropriate for present limited quantum hardware. 

\section*{Acknowledgements}

This work is supported by the National Science Foundation RAISE-QAC-QSA under grant number DMR-2037783 and NSF CAREER Award under Grant No. NSF-ECCS1944085 and the NSF CNS program under Grant No. 2247007.

\appendix  

\section{Details on Implementation with Stochastic Combination of Unitaries}
\label{sec:method}
Focusing on the stochastic combination of unitaries approach, we provide a detailed summary of our approach in Algorithm \eqref{alg:unitary_decomposition}. Implementing the decomposition as described in Eqs.~\eqref{eq:expansionS}-\eqref{eq:expansionA} requires pre-processing the Kraus maps. As discussed in the previous section, we calculate the eigenvalues of the generating Kraus operators and then solve Eq. \eqref{eq:EC=Lambda} to find the expansion coefficients.

\SetKwInput{KwInput}{Input}                
\SetKwInput{KwOutput}{Output}   
\SetKw{KwIn}{in}
\SetKw{KwAnd}{and}
\SetKw{KwSt}{such that} 
\SetKwFunction{Opt}{Optimization-Subroutine}
\SetKwFunction{Init}{Initialize-Params}

\begin{algorithm}[h]
    \DontPrintSemicolon
    \KwInput{$\rho_0, s_{\text{tot}}, \{ M_1, M_2, \ldots, M_k, \ldots \}$}
    \KwOutput{$\rho(t) = \sum_k M_k \rho_0 M_k^\dag$}

    $\mu_{\text{init}} = $ \FuncSty{Initialization}(\ArgSty{$\{ M_1, M_2, \ldots\}$})
    
    \For{$M_k$ \KwIn $\{ M_1, M_2, \ldots \}$}{
        $S, A = \dfrac{1}{2}(M_k + M_k^\dag),~\dfrac{1}{2}(M_k - M_k^\dag)$ \\
        \vspace{6pt}
        $\vec{c}\,{ }^k_{(\text{opt})}, \vec{\mathbf{U}}^k =$ \FuncSty{Optimization-Subroutine}({$\mu_{\text{init}}, S, A$}) \\
        \vspace{6pt}
        $\{s^k_{(i,j)}\} =$ \FuncSty{Multinomial-Sampling}({$s_{\text{tot}}, \vec{c}\,{ }^k_{(\text{opt})}$}) 
        \vspace{6pt}
        
        \For{$c_i^k, c_j^k$ \KwIn $\vec{c}\,{ }^k_{(\text{opt})}$ \KwAnd $U_i^k, U_j^k$ \KwIn $\vec{\mathbf{U}}^k$}{
            $\rho_{(i,j)}^k =$ \FuncSty{Execute}(\ArgSty{$\rho_0, U_i^k, U_j^k, c_i^k, c_j^k, s^k_{(i,j)}$})
        }
        $\rho^k \gets \sum_{i,j}|c_i^k c_j^{k\ast}| \rho_{(i,j)}^k$
        
    }
    $\mathbf{\rho(t) \gets \sum_k \rho^k}$     

    \caption{Exact Unitary Decomposition Simulation}
    \label{alg:unitary_decomposition}
\end{algorithm}

\begin{figure}[ht]
    \begin{quantikz}
        \ket{0}\,& \gate{H} & \gate{P(\alpha)} & \ctrl{1} & \ctrl[open]{1} & \meter{\hat{X}}\\
        \ket{\rho_0}\,& &  & \gate{U_i^k} & \gate{U_j^k}  & \meter{\hat{O}} \, 
    \end{quantikz}
    \caption{This circuit prepares the state $\rho^k_{(i,j)} + \rho^k_{(j,i)}$, and the phase gate applies the phase of $\alpha = c_i^kc_j^{k\ast}$. }
    \label{fig:cross_exec_circ}
\end{figure}
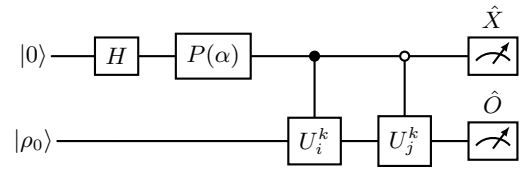

Using the SCU approach, we prepare and stochastically sample unitaries from a normalized distribution according to the $l_1$-cost of a coefficient expansion. The circuits can be implemented with a single ancilla using a circuit akin to a Hadamard test, shown in Fig. \ref{fig:cross_exec_circ}. 

The final density matrix can be written as:
\begin{equation}
    \tilde{\rho} = \sum_k \sum_{i,j} c_{i}^kc_{j}^{k\ast} U_{i}^k \rho U_{j}^{k\dag}
\end{equation}
with coefficients $c_{i}^k \in \{c_{i'}^{k(s)}\} \cup \{c_{i''}\}^{k(a)}$, and unitaries $U_{j}^k \in \{e^{-\iota\mu_{j'}^{k(s)}S_k}\} \cup \{e^{-\mu_{j''}^{k(a)}A_k}\}$, which are assembled in post-processing. Evaluation of the final density matrix involves calculating the self terms ($i=j$) and the cross terms ($i\neq j$).

The total variance can be obtained using standard multinomial circuit sampling \cite{Arrasmith_Cincio_Somma_Coles_2020}. For a given Hermitian involutary observable, the variance with respect to the output state $\tilde{\rho}$ is given as:
\begin{equation}
    \text{Var}(\hat{E}_O) = \dfrac{L^2 - \braket{O}_{\tilde{\rho}}^2}{s_{\text{tot}}}
    \label{eq:variance_kraus}
\end{equation}
where $L = \sum_k\sum_{j,j'}|c_{j'}^{k\ast} c_{j}^k|$ is the $l_1$ norm of \emph{all} expansion coefficients of $\tilde{\rho}$ and $s_{\text{tot}}$ is the total number of shots. While this allows us to reduce the qubit cost over the prior strategy, the variance is asymptotically the same with respect to $\epsilon$.



The maximum number of circuits required to construct the output density matrix scales as $\mathcal{O}(KN^2)$. However, given a budget of total shots, the number of shots assigned to each circuit is dictated by a multinomial distribution. Thus, in practice, we will only sample at most $\mathcal{O}(s_{\rm tot})$ circuits.  

\section{Exact Interpolation for Two Eigenvalues}
\label{sec:interpolation}
Here we demonstrate a solution for the single qubit $N=2$ case which saturates the inequality in Eq.~\eqref{eq:inequal}. We show the derivation for a Hermitian operator, $S$, although this also applies to any anti-Hermitian operator $A$. We can decompose the matrix exponential of $i \mu S$ as:
\begin{equation}
    e^{-\iota\mu_i S} = \Lambda_0^i I + \Lambda_1^i S 
\end{equation}
where
\begin{align}
    \Lambda_0^i &= \dfrac{\lambda_0e^{-\iota\mu_i\lambda_1 } - \lambda_1e^{-\iota\mu_i\lambda_0 }}{\lambda_0 - \lambda_1} \\
    \Lambda_1^i &= \dfrac{e^{-\iota\mu_i\lambda_0 } - e^{-\iota\mu_i\lambda_1 }}{\lambda_0 - \lambda_1} 
\end{align}
are the Lagrangian polynomial coefficients (or Frobenius covariants), and are calculated from the Sylvester matrix theorem. We can construct a linear system of equations:
\begin{equation}
\begin{pmatrix}
    \Lambda_0^0 && \Lambda_0^1 \\    
    \Lambda_1^0 && \Lambda_1^1 
\end{pmatrix}\vec{c} = \begin{pmatrix}
        1 \\ 0
    \end{pmatrix}
\end{equation}
where the left-hand side represents a function evaluated in terms of the coefficients (which are powers of $S$), and the right-hand side yields $S$.  

For a full-rank system the solution is given by $\vec{c}$:
\begin{equation}
\vec{c} = \dfrac{1}{\Delta}\begin{pmatrix}
    -\Lambda^1_0 \\ 
    \Lambda_0^0
\end{pmatrix}
\end{equation}
and $\Delta$ is the determinant of the linear matrix. 

Calculating the $l_1$ norm of the coefficient vector, we find:
\begin{equation}
    |\vec{c}|_1 = \dfrac{|\lambda_0e^{-\iota\mu_1\lambda_1 } - \lambda_1e^{-\iota\mu_1\lambda_0 }| + |\lambda_0e^{-\iota\mu_0\lambda_1 } - \lambda_1e^{-\iota\mu_0\lambda_0 }| }{|e^{-\iota(\mu_0\lambda_1 + \mu_1\lambda_0 )} - e^{-\iota(\mu_0\lambda_0 + \mu_1\lambda_1 )}|}.
\end{equation}
We can simplify this expression assuming $\mu_0 = - \mu_1 = \mu$ and substituting $\lambda_0 = \omega + \sigma$ and $\lambda_1 = \omega - \sigma$
\begin{equation}
    |\vec{c}|_1 = \sqrt{\omega^2\csc^2(\sigma\mu) + \sigma^2\sec^2(\sigma\mu)}.
\end{equation}
We can find the minimal solution corresponding to $\partial_\mu |\vec{c}|_1  = 0$, to get:
\begin{equation}
    \mu_{\min} = \pm\frac{1}{\sigma}\arctan{\bigl(\sqrt{\bigl|\dfrac{\sigma}{\omega}\bigr|} \bigr)}.
\end{equation}
Interestingly, by substituting this into our $l_1$ expression we find that
\begin{equation}
    |\vec{c}|_{1(\min)} = \frac{1}{2}(|\lambda_0 - \lambda_1| + |\lambda_0 + \lambda_1|) = \max{(|\lambda_0|, |\lambda_1|)}
\end{equation}
which saturates the inequality in Eq.~\eqref{eq:inequal}, showing we can obtain an interpolation with $l_1$ norm equivalent to the operator norm.

\section{Detailed Analysis of the Choice of Dilation}
\label{sec:dilations}
Without the interpolation strategy, the SCU can be implemented for the approximate unitary decomposition approach (denoted as AUD), although we show that on average it would has limited advantages over a 2-qubit linear combination of unitaries, mainly a reduction in the number of quantum operators. 

    To show this, we can write the approximate unitary decomposition as:
    \begin{equation}
        \rho = \sum_{\alpha=1}^K\sum_{T T'} \frac{1}{4\epsilon^2} e^{\epsilon T} \rho e^{\epsilon T'^\dagger}
    \end{equation}
    where $T,T'$ are taken from the set of 4 unitaries, and $L = \sum_{\alpha=1}^K |\frac{4}{\epsilon^2} | = \frac{4 K}{\epsilon^2}$. 
with pseudoprobabilities $p_i = \frac{1}{4\epsilon^2}$. As a result, for an involutory observable we have a variance proportional to $\frac{L^2}{s_{\text{tot}}}$ using SCU, which does not provide a substantial benefit over AUD with LCU. 

In the standard LCU approach, we can evaluate the variance for each Kraus term using the law of total variance. Specifically, we obtain:
\begin{equation}
    {\rm Var}_{M_\alpha \rho M_\alpha^\dagger}[A] = p_\alpha - \langle A \rangle^2_{M_\alpha \rho M_\alpha^\dagger} \approx q_\alpha \frac{\epsilon^2}{4}
\end{equation} 
where $p_\alpha$ is the total success probability of the dilation and $q_\alpha$ is the portion specific to the Kraus operators (which forms a proper distribution under all Kraus maps). Note, the total norm $L$ naively is $\frac{4K}{\epsilon^2}$. From this, we can calculate the total variance:
\begin{align}
{\rm Var}_{\tilde{\rho}}[A] &= \frac{4K}{s_{\text{tot}} \epsilon^2} \sum_{\alpha=1}^K \frac{4}{\epsilon^2}{\rm Var}_{M_\alpha \rho M_\alpha^\dagger}[A] \\
&= \frac{4 K}{s_{\text{tot}} \epsilon^2} + \mathcal{O}(\frac{K}{s_{\text{tot}}})
\end{align}
Which is asymptotically different in $\epsilon$ scaling compared with the SCU approach.

Finally, we can perform a standard LCU dilation for the exact approach, albeit it at much larger circuit cost. Letting $\lambda_\alpha$ be the sum of $|\vec{c}^{~A_\alpha}|$ and $|\vec{c}^{~S_\alpha}|$, we have:
\begin{align}
{\rm Var}_{\tilde{\rho}}[A] &= \frac{L}{s_{\rm tot}} \sum_{\alpha=1}^K \lambda_\alpha {\rm Var}_{M_\alpha \rho M_\alpha^\dagger}[A] \\
&= \frac{{L}}{s_{\rm tot}} \sum_{\alpha=1}^K\lambda_\alpha p_\alpha + \mathcal{O}(\frac{{L}}{s_{\rm tot}} p_\alpha^2) \\
&= \mathcal{O}(\frac{{L}}{s_{\rm tot}}) 
\end{align}
Where we use the fact that the Kraus operators can be encoded to have a success probability $\frac{q_\alpha}{\lambda_\alpha}$. We summarize these differences in Table \eqref{tab:scaling}. 

\def\arraystretch{1.75}

\begin{table}[h]
    \centering
\caption{Comparison of variance, (largest) gate cost, and classical computational overhead. $f_c$ represents the average cost of implementing an exponential of $A$ or $S$ with $c$ controls. $n_\alpha$ denotes the number of non-unique eigenvalues averaged in $S_\alpha$ and $A_\alpha$. }
    \begin{tabular}{cc|cc|cc|c}
   &  & \multicolumn{2}{c|}{Variance} & \multicolumn{2}{c|}{Max dilation} & Classical \\
      &  & SCU & LCU & ~SCU~ & LCU &  overhead \\ \hline
    & Exact  
        & $\mathcal{O}(\frac{L^2}{s_{\rm tot}})$ & $\mathcal{O}(\frac{L}{s_{\rm tot}})$ 
            & $2f_1$ & $2n_\alpha f_{\log_2 2 n_\alpha}$ 
                & $\mathcal{O}(\zeta n_\alpha^3)$    \\
    & Approx. 
        & $\mathcal{O}(\frac{K^2}{s_{\rm tot}\epsilon^4})$ & $\mathcal{O}(\frac{K}{s_{\rm tot}\epsilon^2})$ 
            & $2f_1$ & $4f_2$ & $\mathcal{O}(1)$
    \end{tabular}
    \label{tab:scaling}
\end{table}

We see that compared to LCU, SCU yields fewer gates at the cost of larger variances. In practice, this means that SCU requires shorter quantum coherence times but more total samples, $s_{\text{tot}}$, a potentially valuable trade-off. 

When using the approximate approach, we additionally need to consider the error bias from Eq. \eqref{eq:A_expansion_approximate}. These expansions accurately approximate the operator $M$ up to order $\mathcal{O}(\epsilon^2)$. Importantly, limiting this bias to a target precision $E$ thus requires an expansion parameter of at least $\epsilon > \mathcal{O}(\sqrt{E})$, which implies that the total measurement cost of the approximate decomposition under the SCU and LCU dilations have a $\mathcal{O}(E^{-4})$ and $\mathcal{O}(E^{-3})$ measurement cost, respectively. This reflects the fact that due to the bias of the small angle approximation, there is an optimal number of shots for each precision and choice of $\epsilon$.



\begin{thebibliography}{53}%
\makeatletter
\providecommand \@ifxundefined [1]{%
 \@ifx{#1\undefined}
}%
\providecommand \@ifnum [1]{%
 \ifnum #1\expandafter \@firstoftwo
 \else \expandafter \@secondoftwo
 \fi
}%
\providecommand \@ifx [1]{%
 \ifx #1\expandafter \@firstoftwo
 \else \expandafter \@secondoftwo
 \fi
}%
\providecommand \natexlab [1]{#1}%
\providecommand \enquote  [1]{``#1''}%
\providecommand \bibnamefont  [1]{#1}%
\providecommand \bibfnamefont [1]{#1}%
\providecommand \citenamefont [1]{#1}%
\providecommand \href@noop [0]{\@secondoftwo}%
\providecommand \href [0]{\begingroup \@sanitize@url \@href}%
\providecommand \@href[1]{\@@startlink{#1}\@@href}%
\providecommand \@@href[1]{\endgroup#1\@@endlink}%
\providecommand \@sanitize@url [0]{\catcode `\\12\catcode `\$12\catcode
  `\&12\catcode `\#12\catcode `\^12\catcode `\_12\catcode `\%12\relax}%
\providecommand \@@startlink[1]{}%
\providecommand \@@endlink[0]{}%
\providecommand \url  [0]{\begingroup\@sanitize@url \@url }%
\providecommand \@url [1]{\endgroup\@href {#1}{\urlprefix }}%
\providecommand \urlprefix  [0]{URL }%
\providecommand \Eprint [0]{\href }%
\providecommand \doibase [0]{https://doi.org/}%
\providecommand \selectlanguage [0]{\@gobble}%
\providecommand \bibinfo  [0]{\@secondoftwo}%
\providecommand \bibfield  [0]{\@secondoftwo}%
\providecommand \translation [1]{[#1]}%
\providecommand \BibitemOpen [0]{}%
\providecommand \bibitemStop [0]{}%
\providecommand \bibitemNoStop [0]{.\EOS\space}%
\providecommand \EOS [0]{\spacefactor3000\relax}%
\providecommand \BibitemShut  [1]{\csname bibitem#1\endcsname}%
\let\auto@bib@innerbib\@empty
\bibitem [{\citenamefont {Takahashi}(2024)}]{intro_oqs_cmp1}%
  \BibitemOpen
  \bibfield  {author} {\bibinfo {author} {\bibfnamefont {Y.}~\bibnamefont
  {Takahashi}},\ }\bibfield  {title} {\bibinfo {title} {Cold-atom systems as
  condensed matter physics emulation},\ }in\ \href
  {https://doi.org/https://doi.org/10.1016/B978-0-323-90800-9.00271-7} {\emph
  {\bibinfo {booktitle} {Encyclopedia of Condensed Matter Physics (Second
  Edition)}}},\ \bibinfo {editor} {edited by\ \bibinfo {editor} {\bibfnamefont
  {T.}~\bibnamefont {Chakraborty}}}\ (\bibinfo  {publisher} {Academic Press},\
  \bibinfo {address} {Oxford},\ \bibinfo {year} {2024})\ \bibinfo {edition}
  {second edition}\ ed.,\ pp.\ \bibinfo {pages} {135--144}\BibitemShut
  {NoStop}%
\bibitem [{\citenamefont {Okołowicz}\ \emph {et~al.}(2003)\citenamefont
  {Okołowicz}, \citenamefont {Płoszajczak},\ and\ \citenamefont
  {Rotter}}]{intro_oqs_cmp2}%
  \BibitemOpen
  \bibfield  {author} {\bibinfo {author} {\bibfnamefont {J.}~\bibnamefont
  {Okołowicz}}, \bibinfo {author} {\bibfnamefont {M.}~\bibnamefont
  {Płoszajczak}},\ and\ \bibinfo {author} {\bibfnamefont {I.}~\bibnamefont
  {Rotter}},\ }\bibfield  {title} {\bibinfo {title} {Dynamics of quantum
  systems embedded in a continuum},\ }\href
  {https://doi.org/https://doi.org/10.1016/S0370-1573(02)00366-6} {\bibfield
  {journal} {\bibinfo  {journal} {Physics Reports}\ }\textbf {\bibinfo {volume}
  {374}},\ \bibinfo {pages} {271} (\bibinfo {year} {2003})}\BibitemShut
  {NoStop}%
\bibitem [{\citenamefont {Akamatsu}(2022)}]{intro_oqs_cmp3}%
  \BibitemOpen
  \bibfield  {author} {\bibinfo {author} {\bibfnamefont {Y.}~\bibnamefont
  {Akamatsu}},\ }\bibfield  {title} {\bibinfo {title} {Quarkonium in
  quark–gluon plasma: Open quantum system approaches re-examined},\ }\href
  {https://doi.org/https://doi.org/10.1016/j.ppnp.2021.103932} {\bibfield
  {journal} {\bibinfo  {journal} {Progress in Particle and Nuclear Physics}\
  }\textbf {\bibinfo {volume} {123}},\ \bibinfo {pages} {103932} (\bibinfo
  {year} {2022})}\BibitemShut {NoStop}%
\bibitem [{\citenamefont {Roduner}\ and\ \citenamefont
  {Krüger}(2022)}]{intro_oqs_cmp4_thermo}%
  \BibitemOpen
  \bibfield  {author} {\bibinfo {author} {\bibfnamefont {E.}~\bibnamefont
  {Roduner}}\ and\ \bibinfo {author} {\bibfnamefont {T.~P.}\ \bibnamefont
  {Krüger}},\ }\bibfield  {title} {\bibinfo {title} {The origin of
  irreversibility and thermalization in thermodynamic processes},\ }\href
  {https://doi.org/https://doi.org/10.1016/j.physrep.2021.11.002} {\bibfield
  {journal} {\bibinfo  {journal} {Physics Reports}\ }\textbf {\bibinfo {volume}
  {944}},\ \bibinfo {pages} {1} (\bibinfo {year} {2022})},\ \bibinfo {note}
  {the origin of irreversibility and thermalization in thermodynamic
  processes}\BibitemShut {NoStop}%
\bibitem [{\citenamefont {Delle Site}\ and\ \citenamefont
  {Praprotnik}(2017)}]{intro_oqs_chem1}%
  \BibitemOpen
  \bibfield  {author} {\bibinfo {author} {\bibfnamefont {L.}~\bibnamefont
  {Delle Site}}\ and\ \bibinfo {author} {\bibfnamefont {M.}~\bibnamefont
  {Praprotnik}},\ }\bibfield  {title} {\bibinfo {title} {Molecular systems with
  open boundaries: Theory and simulation},\ }\href
  {https://doi.org/https://doi.org/10.1016/j.physrep.2017.05.007} {\bibfield
  {journal} {\bibinfo  {journal} {Physics Reports}\ }\textbf {\bibinfo {volume}
  {693}},\ \bibinfo {pages} {1} (\bibinfo {year} {2017})},\ \bibinfo {note}
  {molecular systems with open boundaries: Theory and Simulation}\BibitemShut
  {NoStop}%
\bibitem [{\citenamefont {Yunger~Halpern}\ \emph {et~al.}(2016)\citenamefont
  {Yunger~Halpern}, \citenamefont {Faist}, \citenamefont {Oppenheim},\ and\
  \citenamefont {Winter}}]{intro_oqs_thermo2}%
  \BibitemOpen
  \bibfield  {author} {\bibinfo {author} {\bibfnamefont {N.}~\bibnamefont
  {Yunger~Halpern}}, \bibinfo {author} {\bibfnamefont {P.}~\bibnamefont
  {Faist}}, \bibinfo {author} {\bibfnamefont {J.}~\bibnamefont {Oppenheim}},\
  and\ \bibinfo {author} {\bibfnamefont {A.}~\bibnamefont {Winter}},\
  }\bibfield  {title} {\bibinfo {title} {Microcanonical and resource-theoretic
  derivations of the thermal state of a quantum system with noncommuting
  charges},\ }\bibfield  {journal} {\bibinfo  {journal} {Nature
  Communications}\ }\textbf {\bibinfo {volume} {7}},\ \href
  {https://doi.org/10.1038/ncomms12051} {10.1038/ncomms12051} (\bibinfo {year}
  {2016})\BibitemShut {NoStop}%
\bibitem [{\citenamefont {Martin}\ \emph {et~al.}(2013)\citenamefont {Martin},
  \citenamefont {Bishof}, \citenamefont {Swallows}, \citenamefont {Zhang},
  \citenamefont {Benko}, \citenamefont {von Stecher}, \citenamefont {Gorshkov},
  \citenamefont {Rey},\ and\ \citenamefont {Ye}}]{intro_oqs_metro1}%
  \BibitemOpen
  \bibfield  {author} {\bibinfo {author} {\bibfnamefont {M.~J.}\ \bibnamefont
  {Martin}}, \bibinfo {author} {\bibfnamefont {M.}~\bibnamefont {Bishof}},
  \bibinfo {author} {\bibfnamefont {M.~D.}\ \bibnamefont {Swallows}}, \bibinfo
  {author} {\bibfnamefont {X.}~\bibnamefont {Zhang}}, \bibinfo {author}
  {\bibfnamefont {C.}~\bibnamefont {Benko}}, \bibinfo {author} {\bibfnamefont
  {J.}~\bibnamefont {von Stecher}}, \bibinfo {author} {\bibfnamefont {A.~V.}\
  \bibnamefont {Gorshkov}}, \bibinfo {author} {\bibfnamefont {A.~M.}\
  \bibnamefont {Rey}},\ and\ \bibinfo {author} {\bibfnamefont {J.}~\bibnamefont
  {Ye}},\ }\bibfield  {title} {\bibinfo {title} {A quantum many-body spin
  system in an optical lattice clock},\ }\href
  {https://doi.org/10.1126/science.1236929} {\bibfield  {journal} {\bibinfo
  {journal} {Science}\ }\textbf {\bibinfo {volume} {341}},\ \bibinfo {pages}
  {632} (\bibinfo {year} {2013})},\ \Eprint
  {https://arxiv.org/abs/https://www.science.org/doi/pdf/10.1126/science.1236929}
  {https://www.science.org/doi/pdf/10.1126/science.1236929} \BibitemShut
  {NoStop}%
\bibitem [{\citenamefont {Kolkowitz}\ \emph {et~al.}(2017)\citenamefont
  {Kolkowitz}, \citenamefont {Bromley}, \citenamefont {Bothwell}, \citenamefont
  {Wall}, \citenamefont {Marti}, \citenamefont {Koller}, \citenamefont {Zhang},
  \citenamefont {Rey},\ and\ \citenamefont {Ye}}]{intro_oqs_metro2}%
  \BibitemOpen
  \bibfield  {author} {\bibinfo {author} {\bibfnamefont {S.}~\bibnamefont
  {Kolkowitz}}, \bibinfo {author} {\bibfnamefont {S.~L.}\ \bibnamefont
  {Bromley}}, \bibinfo {author} {\bibfnamefont {T.}~\bibnamefont {Bothwell}},
  \bibinfo {author} {\bibfnamefont {M.~L.}\ \bibnamefont {Wall}}, \bibinfo
  {author} {\bibfnamefont {G.~E.}\ \bibnamefont {Marti}}, \bibinfo {author}
  {\bibfnamefont {A.~P.}\ \bibnamefont {Koller}}, \bibinfo {author}
  {\bibfnamefont {X.}~\bibnamefont {Zhang}}, \bibinfo {author} {\bibfnamefont
  {A.~M.}\ \bibnamefont {Rey}},\ and\ \bibinfo {author} {\bibfnamefont
  {J.}~\bibnamefont {Ye}},\ }\bibfield  {title} {\bibinfo {title}
  {Spin–orbit-coupled fermions in an optical lattice clock},\ }\href
  {https://doi.org/10.1038/nature20811} {\bibfield  {journal} {\bibinfo
  {journal} {Nature}\ }\textbf {\bibinfo {volume} {542}},\ \bibinfo {pages}
  {66–70} (\bibinfo {year} {2017})}\BibitemShut {NoStop}%
\bibitem [{\citenamefont {Kraus}\ \emph {et~al.}(2008)\citenamefont {Kraus},
  \citenamefont {B\"uchler}, \citenamefont {Diehl}, \citenamefont {Kantian},
  \citenamefont {Micheli},\ and\ \citenamefont {Zoller}}]{intro_oqs_qis1}%
  \BibitemOpen
  \bibfield  {author} {\bibinfo {author} {\bibfnamefont {B.}~\bibnamefont
  {Kraus}}, \bibinfo {author} {\bibfnamefont {H.~P.}\ \bibnamefont
  {B\"uchler}}, \bibinfo {author} {\bibfnamefont {S.}~\bibnamefont {Diehl}},
  \bibinfo {author} {\bibfnamefont {A.}~\bibnamefont {Kantian}}, \bibinfo
  {author} {\bibfnamefont {A.}~\bibnamefont {Micheli}},\ and\ \bibinfo {author}
  {\bibfnamefont {P.}~\bibnamefont {Zoller}},\ }\bibfield  {title} {\bibinfo
  {title} {Preparation of entangled states by quantum markov processes},\
  }\href {https://doi.org/10.1103/PhysRevA.78.042307} {\bibfield  {journal}
  {\bibinfo  {journal} {Phys. Rev. A}\ }\textbf {\bibinfo {volume} {78}},\
  \bibinfo {pages} {042307} (\bibinfo {year} {2008})}\BibitemShut {NoStop}%
\bibitem [{\citenamefont {Diehl}\ \emph {et~al.}(2008)\citenamefont {Diehl},
  \citenamefont {Micheli}, \citenamefont {Kantian}, \citenamefont {Kraus},
  \citenamefont {Büchler},\ and\ \citenamefont {Zoller}}]{intro_oqs_qis2}%
  \BibitemOpen
  \bibfield  {author} {\bibinfo {author} {\bibfnamefont {S.}~\bibnamefont
  {Diehl}}, \bibinfo {author} {\bibfnamefont {A.}~\bibnamefont {Micheli}},
  \bibinfo {author} {\bibfnamefont {A.}~\bibnamefont {Kantian}}, \bibinfo
  {author} {\bibfnamefont {B.}~\bibnamefont {Kraus}}, \bibinfo {author}
  {\bibfnamefont {H.~P.}\ \bibnamefont {Büchler}},\ and\ \bibinfo {author}
  {\bibfnamefont {P.}~\bibnamefont {Zoller}},\ }\bibfield  {title} {\bibinfo
  {title} {Quantum states and phases in driven open quantum systems with cold
  atoms},\ }\href {https://doi.org/10.1038/nphys1073} {\bibfield  {journal}
  {\bibinfo  {journal} {Nature Physics}\ }\textbf {\bibinfo {volume} {4}},\
  \bibinfo {pages} {878–883} (\bibinfo {year} {2008})}\BibitemShut {NoStop}%
\bibitem [{\citenamefont {Head-Marsden}\ \emph
  {et~al.}(2021{\natexlab{a}})\citenamefont {Head-Marsden}, \citenamefont
  {Flick}, \citenamefont {Ciccarino},\ and\ \citenamefont
  {Narang}}]{intro_oqs_qis3}%
  \BibitemOpen
  \bibfield  {author} {\bibinfo {author} {\bibfnamefont {K.}~\bibnamefont
  {Head-Marsden}}, \bibinfo {author} {\bibfnamefont {J.}~\bibnamefont {Flick}},
  \bibinfo {author} {\bibfnamefont {C.~J.}\ \bibnamefont {Ciccarino}},\ and\
  \bibinfo {author} {\bibfnamefont {P.}~\bibnamefont {Narang}},\ }\bibfield
  {title} {\bibinfo {title} {Quantum information and algorithms for correlated
  quantum matter},\ }\href {https://doi.org/10.1021/acs.chemrev.0c00620}
  {\bibfield  {journal} {\bibinfo  {journal} {Chemical Reviews}\ }\textbf
  {\bibinfo {volume} {121}},\ \bibinfo {pages} {3061–3120} (\bibinfo {year}
  {2021}{\natexlab{a}})}\BibitemShut {NoStop}%
\bibitem [{\citenamefont {Olivera‐Atencio}\ \emph {et~al.}(2023)\citenamefont
  {Olivera‐Atencio}, \citenamefont {Lamata},\ and\ \citenamefont
  {Casado‐Pascual}}]{intro_oqs_qis4}%
  \BibitemOpen
  \bibfield  {author} {\bibinfo {author} {\bibfnamefont {M.~L.}\ \bibnamefont
  {Olivera‐Atencio}}, \bibinfo {author} {\bibfnamefont {L.}~\bibnamefont
  {Lamata}},\ and\ \bibinfo {author} {\bibfnamefont {J.}~\bibnamefont
  {Casado‐Pascual}},\ }\bibfield  {title} {\bibinfo {title} {Benefits of open
  quantum systems for quantum machine learning},\ }\href
  {https://doi.org/10.1002/qute.202300247} {\bibfield  {journal} {\bibinfo
  {journal} {Advanced Quantum Technologies}\ ,\ \bibinfo {pages} {2300247}}
  (\bibinfo {year} {2023})}\BibitemShut {NoStop}%
\bibitem [{\citenamefont {Georgescu}\ \emph {et~al.}(2014)\citenamefont
  {Georgescu}, \citenamefont {Ashhab},\ and\ \citenamefont
  {Nori}}]{intro_oqs_qis5}%
  \BibitemOpen
  \bibfield  {author} {\bibinfo {author} {\bibfnamefont {I.~M.}\ \bibnamefont
  {Georgescu}}, \bibinfo {author} {\bibfnamefont {S.}~\bibnamefont {Ashhab}},\
  and\ \bibinfo {author} {\bibfnamefont {F.}~\bibnamefont {Nori}},\ }\bibfield
  {title} {\bibinfo {title} {Quantum simulation},\ }\href
  {https://doi.org/10.1103/RevModPhys.86.153} {\bibfield  {journal} {\bibinfo
  {journal} {Rev. Mod. Phys.}\ }\textbf {\bibinfo {volume} {86}},\ \bibinfo
  {pages} {153} (\bibinfo {year} {2014})}\BibitemShut {NoStop}%
\bibitem [{\citenamefont {Breuer}\ and\ \citenamefont
  {Petruccione}(2007)}]{theory_oqs_book}%
  \BibitemOpen
  \bibfield  {author} {\bibinfo {author} {\bibfnamefont {H.-P.}\ \bibnamefont
  {Breuer}}\ and\ \bibinfo {author} {\bibfnamefont {F.}~\bibnamefont
  {Petruccione}},\ }\href@noop {} {\emph {\bibinfo {title} {The theory of Open
  Quantum Systems}}}\ (\bibinfo  {publisher} {Clarendon},\ \bibinfo {year}
  {2007})\BibitemShut {NoStop}%
\bibitem [{\citenamefont {Lindblad}(1976)}]{lindblad_master_eq}%
  \BibitemOpen
  \bibfield  {author} {\bibinfo {author} {\bibfnamefont {G.}~\bibnamefont
  {Lindblad}},\ }\bibfield  {title} {\bibinfo {title} {On the generators of
  quantum dynamical semigroups},\ }\href {https://doi.org/10.1007/bf01608499}
  {\bibfield  {journal} {\bibinfo  {journal} {Communications in Mathematical
  Physics}\ }\textbf {\bibinfo {volume} {48}},\ \bibinfo {pages} {119–130}
  (\bibinfo {year} {1976})}\BibitemShut {NoStop}%
\bibitem [{\citenamefont {Gorini}\ \emph {et~al.}(1976)\citenamefont {Gorini},
  \citenamefont {Kossakowski},\ and\ \citenamefont {Sudarshan}}]{cpds_oqs}%
  \BibitemOpen
  \bibfield  {author} {\bibinfo {author} {\bibfnamefont {V.}~\bibnamefont
  {Gorini}}, \bibinfo {author} {\bibfnamefont {A.}~\bibnamefont
  {Kossakowski}},\ and\ \bibinfo {author} {\bibfnamefont {E.~C.}\ \bibnamefont
  {Sudarshan}},\ }\bibfield  {title} {\bibinfo {title} {Completely positive
  dynamical semigroups of n-level systems},\ }\href
  {https://doi.org/10.1063/1.522979} {\bibfield  {journal} {\bibinfo  {journal}
  {Journal of Mathematical Physics}\ }\textbf {\bibinfo {volume} {17}},\
  \bibinfo {pages} {821–825} (\bibinfo {year} {1976})}\BibitemShut {NoStop}%
\bibitem [{\citenamefont {Stinespring}(1955)}]{Stinespring_dilation}%
  \BibitemOpen
  \bibfield  {author} {\bibinfo {author} {\bibfnamefont {W.~F.}\ \bibnamefont
  {Stinespring}},\ }\bibfield  {title} {\bibinfo {title} {Positive functions on
  c*-algebras},\ }\href {https://doi.org/10.1090/s0002-9939-1955-0069403-4}
  {\bibfield  {journal} {\bibinfo  {journal} {Proceedings of the American
  Mathematical Society}\ }\textbf {\bibinfo {volume} {6}},\ \bibinfo {pages}
  {211–216} (\bibinfo {year} {1955})}\BibitemShut {NoStop}%
\bibitem [{\citenamefont {Paulsen}(2003)}]{dilation_paulsen}%
  \BibitemOpen
  \bibfield  {author} {\bibinfo {author} {\bibfnamefont {V.}~\bibnamefont
  {Paulsen}},\ }\href@noop {} {\emph {\bibinfo {title} {Completely Bounded Maps
  and Operator Algebras}}},\ Cambridge Studies in Advanced Mathematics\
  (\bibinfo  {publisher} {Cambridge University Press},\ \bibinfo {year}
  {2003})\BibitemShut {NoStop}%
\bibitem [{\citenamefont {Langer}(1972)}]{dilation_sz_nagy}%
  \BibitemOpen
  \bibfield  {author} {\bibinfo {author} {\bibfnamefont {H.}~\bibnamefont
  {Langer}},\ }\bibfield  {title} {\bibinfo {title} {B. sz.‐nagy and c.
  foias, harmonic analysis of operators on hilbert space. viii + 387 s.
  budapest/amsterdam/london 1970. akadémiai kiadó/north‐holland publishing
  company},\ }\href {https://doi.org/10.1002/zamm.19720520821} {\bibfield
  {journal} {\bibinfo  {journal} {ZAMM - Journal of Applied Mathematics and
  Mechanics / Zeitschrift für Angewandte Mathematik und Mechanik}\ }\textbf
  {\bibinfo {volume} {52}},\ \bibinfo {pages} {501–501} (\bibinfo {year}
  {1972})}\BibitemShut {NoStop}%
\bibitem [{\citenamefont {Kliesch}\ \emph {et~al.}(2011)\citenamefont
  {Kliesch}, \citenamefont {Barthel}, \citenamefont {Gogolin}, \citenamefont
  {Kastoryano},\ and\ \citenamefont {Eisert}}]{lindbladian_trotter1}%
  \BibitemOpen
  \bibfield  {author} {\bibinfo {author} {\bibfnamefont {M.}~\bibnamefont
  {Kliesch}}, \bibinfo {author} {\bibfnamefont {T.}~\bibnamefont {Barthel}},
  \bibinfo {author} {\bibfnamefont {C.}~\bibnamefont {Gogolin}}, \bibinfo
  {author} {\bibfnamefont {M.}~\bibnamefont {Kastoryano}},\ and\ \bibinfo
  {author} {\bibfnamefont {J.}~\bibnamefont {Eisert}},\ }\bibfield  {title}
  {\bibinfo {title} {Dissipative quantum church-turing theorem},\ }\href
  {https://doi.org/10.1103/PhysRevLett.107.120501} {\bibfield  {journal}
  {\bibinfo  {journal} {Phys. Rev. Lett.}\ }\textbf {\bibinfo {volume} {107}},\
  \bibinfo {pages} {120501} (\bibinfo {year} {2011})}\BibitemShut {NoStop}%
\bibitem [{\citenamefont {Wang}\ \emph {et~al.}(2011)\citenamefont {Wang},
  \citenamefont {Ashhab},\ and\ \citenamefont {Nori}}]{lindbladian_trotter2}%
  \BibitemOpen
  \bibfield  {author} {\bibinfo {author} {\bibfnamefont {H.}~\bibnamefont
  {Wang}}, \bibinfo {author} {\bibfnamefont {S.}~\bibnamefont {Ashhab}},\ and\
  \bibinfo {author} {\bibfnamefont {F.}~\bibnamefont {Nori}},\ }\bibfield
  {title} {\bibinfo {title} {Quantum algorithm for simulating the dynamics of
  an open quantum system},\ }\href {https://doi.org/10.1103/PhysRevA.83.062317}
  {\bibfield  {journal} {\bibinfo  {journal} {Phys. Rev. A}\ }\textbf {\bibinfo
  {volume} {83}},\ \bibinfo {pages} {062317} (\bibinfo {year}
  {2011})}\BibitemShut {NoStop}%
\bibitem [{\citenamefont {Barthel}\ and\ \citenamefont
  {Kliesch}(2012)}]{lindbladian_trotter3}%
  \BibitemOpen
  \bibfield  {author} {\bibinfo {author} {\bibfnamefont {T.}~\bibnamefont
  {Barthel}}\ and\ \bibinfo {author} {\bibfnamefont {M.}~\bibnamefont
  {Kliesch}},\ }\bibfield  {title} {\bibinfo {title} {Quasilocality and
  efficient simulation of markovian quantum dynamics},\ }\href
  {https://doi.org/10.1103/PhysRevLett.108.230504} {\bibfield  {journal}
  {\bibinfo  {journal} {Phys. Rev. Lett.}\ }\textbf {\bibinfo {volume} {108}},\
  \bibinfo {pages} {230504} (\bibinfo {year} {2012})}\BibitemShut {NoStop}%
\bibitem [{\citenamefont {Han}\ \emph {et~al.}(2021)\citenamefont {Han},
  \citenamefont {Cai}, \citenamefont {Hu}, \citenamefont {Mu}, \citenamefont
  {Ma}, \citenamefont {Xu}, \citenamefont {Wang}, \citenamefont {Wang},
  \citenamefont {Song}, \citenamefont {Zou},\ and\ \citenamefont
  {Sun}}]{lindbladian_trotter4}%
  \BibitemOpen
  \bibfield  {author} {\bibinfo {author} {\bibfnamefont {J.}~\bibnamefont
  {Han}}, \bibinfo {author} {\bibfnamefont {W.}~\bibnamefont {Cai}}, \bibinfo
  {author} {\bibfnamefont {L.}~\bibnamefont {Hu}}, \bibinfo {author}
  {\bibfnamefont {X.}~\bibnamefont {Mu}}, \bibinfo {author} {\bibfnamefont
  {Y.}~\bibnamefont {Ma}}, \bibinfo {author} {\bibfnamefont {Y.}~\bibnamefont
  {Xu}}, \bibinfo {author} {\bibfnamefont {W.}~\bibnamefont {Wang}}, \bibinfo
  {author} {\bibfnamefont {H.}~\bibnamefont {Wang}}, \bibinfo {author}
  {\bibfnamefont {Y.~P.}\ \bibnamefont {Song}}, \bibinfo {author}
  {\bibfnamefont {C.-L.}\ \bibnamefont {Zou}},\ and\ \bibinfo {author}
  {\bibfnamefont {L.}~\bibnamefont {Sun}},\ }\bibfield  {title} {\bibinfo
  {title} {Experimental simulation of open quantum system dynamics via
  trotterization},\ }\href {https://doi.org/10.1103/PhysRevLett.127.020504}
  {\bibfield  {journal} {\bibinfo  {journal} {Phys. Rev. Lett.}\ }\textbf
  {\bibinfo {volume} {127}},\ \bibinfo {pages} {020504} (\bibinfo {year}
  {2021})}\BibitemShut {NoStop}%
\bibitem [{\citenamefont {Bacon}\ \emph {et~al.}(2001)\citenamefont {Bacon},
  \citenamefont {Childs}, \citenamefont {Chuang}, \citenamefont {Kempe},
  \citenamefont {Leung},\ and\ \citenamefont {Zhou}}]{lindbladian_sim1}%
  \BibitemOpen
  \bibfield  {author} {\bibinfo {author} {\bibfnamefont {D.}~\bibnamefont
  {Bacon}}, \bibinfo {author} {\bibfnamefont {A.~M.}\ \bibnamefont {Childs}},
  \bibinfo {author} {\bibfnamefont {I.~L.}\ \bibnamefont {Chuang}}, \bibinfo
  {author} {\bibfnamefont {J.}~\bibnamefont {Kempe}}, \bibinfo {author}
  {\bibfnamefont {D.~W.}\ \bibnamefont {Leung}},\ and\ \bibinfo {author}
  {\bibfnamefont {X.}~\bibnamefont {Zhou}},\ }\bibfield  {title} {\bibinfo
  {title} {Universal simulation of markovian quantum dynamics},\ }\href
  {https://doi.org/10.1103/PhysRevA.64.062302} {\bibfield  {journal} {\bibinfo
  {journal} {Phys. Rev. A}\ }\textbf {\bibinfo {volume} {64}},\ \bibinfo
  {pages} {062302} (\bibinfo {year} {2001})}\BibitemShut {NoStop}%
\bibitem [{\citenamefont {Sweke}\ \emph {et~al.}(2015)\citenamefont {Sweke},
  \citenamefont {Sinayskiy}, \citenamefont {Bernard},\ and\ \citenamefont
  {Petruccione}}]{lindbladian_sim2}%
  \BibitemOpen
  \bibfield  {author} {\bibinfo {author} {\bibfnamefont {R.}~\bibnamefont
  {Sweke}}, \bibinfo {author} {\bibfnamefont {I.}~\bibnamefont {Sinayskiy}},
  \bibinfo {author} {\bibfnamefont {D.}~\bibnamefont {Bernard}},\ and\ \bibinfo
  {author} {\bibfnamefont {F.}~\bibnamefont {Petruccione}},\ }\bibfield
  {title} {\bibinfo {title} {Universal simulation of markovian open quantum
  systems},\ }\href {https://doi.org/10.1103/PhysRevA.91.062308} {\bibfield
  {journal} {\bibinfo  {journal} {Phys. Rev. A}\ }\textbf {\bibinfo {volume}
  {91}},\ \bibinfo {pages} {062308} (\bibinfo {year} {2015})}\BibitemShut
  {NoStop}%
\bibitem [{\citenamefont {Cleve}\ and\ \citenamefont
  {Wang}(2017)}]{lindbladian_efficient_simulation}%
  \BibitemOpen
  \bibfield  {author} {\bibinfo {author} {\bibfnamefont {R.}~\bibnamefont
  {Cleve}}\ and\ \bibinfo {author} {\bibfnamefont {C.}~\bibnamefont {Wang}},\
  }\bibfield  {title} {\bibinfo {title} {{Efficient Quantum Algorithms for
  Simulating Lindblad Evolution}},\ }in\ \href
  {https://doi.org/10.4230/LIPIcs.ICALP.2017.17} {\emph {\bibinfo {booktitle}
  {44th International Colloquium on Automata, Languages, and Programming (ICALP
  2017)}}},\ \bibinfo {series} {Leibniz International Proceedings in
  Informatics (LIPIcs)}, Vol.~\bibinfo {volume} {80},\ \bibinfo {editor}
  {edited by\ \bibinfo {editor} {\bibfnamefont {I.}~\bibnamefont
  {Chatzigiannakis}}, \bibinfo {editor} {\bibfnamefont {P.}~\bibnamefont
  {Indyk}}, \bibinfo {editor} {\bibfnamefont {F.}~\bibnamefont {Kuhn}},\ and\
  \bibinfo {editor} {\bibfnamefont {A.}~\bibnamefont {Muscholl}}}\ (\bibinfo
  {publisher} {Schloss Dagstuhl -- Leibniz-Zentrum f{\"u}r Informatik},\
  \bibinfo {address} {Dagstuhl, Germany},\ \bibinfo {year} {2017})\ pp.\
  \bibinfo {pages} {17:1--17:14}\BibitemShut {NoStop}%
\bibitem [{\citenamefont {Hu}\ \emph {et~al.}(2020)\citenamefont {Hu},
  \citenamefont {Xia},\ and\ \citenamefont {Kais}}]{intro_sznagy_1}%
  \BibitemOpen
  \bibfield  {author} {\bibinfo {author} {\bibfnamefont {Z.}~\bibnamefont
  {Hu}}, \bibinfo {author} {\bibfnamefont {R.}~\bibnamefont {Xia}},\ and\
  \bibinfo {author} {\bibfnamefont {S.}~\bibnamefont {Kais}},\ }\bibfield
  {title} {\bibinfo {title} {A quantum algorithm for evolving open quantum
  dynamics on quantum computing devices},\ }\bibfield  {journal} {\bibinfo
  {journal} {Scientific Reports}\ }\textbf {\bibinfo {volume} {10}},\ \href
  {https://doi.org/10.1038/s41598-020-60321-x} {10.1038/s41598-020-60321-x}
  (\bibinfo {year} {2020})\BibitemShut {NoStop}%
\bibitem [{\citenamefont {Walters}\ and\ \citenamefont
  {Wang}(2024)}]{intro_sznagy_2}%
  \BibitemOpen
  \bibfield  {author} {\bibinfo {author} {\bibfnamefont {P.~L.}\ \bibnamefont
  {Walters}}\ and\ \bibinfo {author} {\bibfnamefont {F.}~\bibnamefont {Wang}},\
  }\bibfield  {title} {\bibinfo {title} {Path integral quantum algorithm for
  simulating non-markovian quantum dynamics in open quantum systems},\
  }\bibfield  {journal} {\bibinfo  {journal} {Physical Review Research}\
  }\textbf {\bibinfo {volume} {6}},\ \href
  {https://doi.org/10.1103/physrevresearch.6.013135}
  {10.1103/physrevresearch.6.013135} (\bibinfo {year} {2024})\BibitemShut
  {NoStop}%
\bibitem [{\citenamefont {Wang}\ \emph {et~al.}(2023)\citenamefont {Wang},
  \citenamefont {Mulvihill}, \citenamefont {Hu}, \citenamefont {Lyu},
  \citenamefont {Shivpuje}, \citenamefont {Liu}, \citenamefont {Soley},
  \citenamefont {Geva}, \citenamefont {Batista},\ and\ \citenamefont
  {Kais}}]{intro_sznagy_3}%
  \BibitemOpen
  \bibfield  {author} {\bibinfo {author} {\bibfnamefont {Y.}~\bibnamefont
  {Wang}}, \bibinfo {author} {\bibfnamefont {E.}~\bibnamefont {Mulvihill}},
  \bibinfo {author} {\bibfnamefont {Z.}~\bibnamefont {Hu}}, \bibinfo {author}
  {\bibfnamefont {N.}~\bibnamefont {Lyu}}, \bibinfo {author} {\bibfnamefont
  {S.}~\bibnamefont {Shivpuje}}, \bibinfo {author} {\bibfnamefont
  {Y.}~\bibnamefont {Liu}}, \bibinfo {author} {\bibfnamefont {M.~B.}\
  \bibnamefont {Soley}}, \bibinfo {author} {\bibfnamefont {E.}~\bibnamefont
  {Geva}}, \bibinfo {author} {\bibfnamefont {V.~S.}\ \bibnamefont {Batista}},\
  and\ \bibinfo {author} {\bibfnamefont {S.}~\bibnamefont {Kais}},\ }\bibfield
  {title} {\bibinfo {title} {Simulating open quantum system dynamics on nisq
  computers with generalized quantum master equations},\ }\href
  {https://doi.org/10.1021/acs.jctc.3c00316} {\bibfield  {journal} {\bibinfo
  {journal} {Journal of Chemical Theory and Computation}\ }\textbf {\bibinfo
  {volume} {19}},\ \bibinfo {pages} {4851–4862} (\bibinfo {year}
  {2023})}\BibitemShut {NoStop}%
\bibitem [{\citenamefont {Gaikwad}\ \emph {et~al.}(2022)\citenamefont
  {Gaikwad}, \citenamefont {Arvind},\ and\ \citenamefont
  {Dorai}}]{sz-nagy_nmr}%
  \BibitemOpen
  \bibfield  {author} {\bibinfo {author} {\bibfnamefont {A.}~\bibnamefont
  {Gaikwad}}, \bibinfo {author} {\bibnamefont {Arvind}},\ and\ \bibinfo
  {author} {\bibfnamefont {K.}~\bibnamefont {Dorai}},\ }\bibfield  {title}
  {\bibinfo {title} {Simulating open quantum dynamics on an nmr quantum
  processor using the sz.-nagy dilation algorithm},\ }\href
  {https://doi.org/10.1103/PhysRevA.106.022424} {\bibfield  {journal} {\bibinfo
   {journal} {Phys. Rev. A}\ }\textbf {\bibinfo {volume} {106}},\ \bibinfo
  {pages} {022424} (\bibinfo {year} {2022})}\BibitemShut {NoStop}%
\bibitem [{\citenamefont {Hu}\ \emph {et~al.}(2022)\citenamefont {Hu},
  \citenamefont {Head-Marsden}, \citenamefont {Mazziotti}, \citenamefont
  {Narang},\ and\ \citenamefont {Kais}}]{sz-nagy_fmo_complex}%
  \BibitemOpen
  \bibfield  {author} {\bibinfo {author} {\bibfnamefont {Z.}~\bibnamefont
  {Hu}}, \bibinfo {author} {\bibfnamefont {K.}~\bibnamefont {Head-Marsden}},
  \bibinfo {author} {\bibfnamefont {D.~A.}\ \bibnamefont {Mazziotti}}, \bibinfo
  {author} {\bibfnamefont {P.}~\bibnamefont {Narang}},\ and\ \bibinfo {author}
  {\bibfnamefont {S.}~\bibnamefont {Kais}},\ }\bibfield  {title} {\bibinfo
  {title} {A general quantum algorithm for open quantum dynamics demonstrated
  with the fenna-matthews-olson complex},\ }\href
  {https://doi.org/10.22331/q-2022-05-30-726} {\bibfield  {journal} {\bibinfo
  {journal} {Quantum}\ }\textbf {\bibinfo {volume} {6}},\ \bibinfo {pages}
  {726} (\bibinfo {year} {2022})}\BibitemShut {NoStop}%
\bibitem [{\citenamefont {Head-Marsden}\ \emph
  {et~al.}(2021{\natexlab{b}})\citenamefont {Head-Marsden}, \citenamefont
  {Krastanov}, \citenamefont {Mazziotti},\ and\ \citenamefont
  {Narang}}]{sz-nagy_david-pri}%
  \BibitemOpen
  \bibfield  {author} {\bibinfo {author} {\bibfnamefont {K.}~\bibnamefont
  {Head-Marsden}}, \bibinfo {author} {\bibfnamefont {S.}~\bibnamefont
  {Krastanov}}, \bibinfo {author} {\bibfnamefont {D.~A.}\ \bibnamefont
  {Mazziotti}},\ and\ \bibinfo {author} {\bibfnamefont {P.}~\bibnamefont
  {Narang}},\ }\bibfield  {title} {\bibinfo {title} {Capturing non-markovian
  dynamics on near-term quantum computers},\ }\href
  {https://doi.org/10.1103/PhysRevResearch.3.013182} {\bibfield  {journal}
  {\bibinfo  {journal} {Phys. Rev. Res.}\ }\textbf {\bibinfo {volume} {3}},\
  \bibinfo {pages} {013182} (\bibinfo {year} {2021}{\natexlab{b}})}\BibitemShut
  {NoStop}%
\bibitem [{\citenamefont {Seneviratne}\ \emph {et~al.}(2024)\citenamefont
  {Seneviratne}, \citenamefont {Walters},\ and\ \citenamefont
  {Wang}}]{intro_svd_1}%
  \BibitemOpen
  \bibfield  {author} {\bibinfo {author} {\bibfnamefont {A.}~\bibnamefont
  {Seneviratne}}, \bibinfo {author} {\bibfnamefont {P.~L.}\ \bibnamefont
  {Walters}},\ and\ \bibinfo {author} {\bibfnamefont {F.}~\bibnamefont
  {Wang}},\ }\bibfield  {title} {\bibinfo {title} {Exact non-markovian quantum
  dynamics on the nisq device using kraus operators},\ }\href
  {https://doi.org/10.1021/acsomega.3c09720} {\bibfield  {journal} {\bibinfo
  {journal} {ACS Omega}\ }\textbf {\bibinfo {volume} {9}},\ \bibinfo {pages}
  {9666–9675} (\bibinfo {year} {2024})}\BibitemShut {NoStop}%
\bibitem [{\citenamefont {Dan}\ \emph {et~al.}(2024)\citenamefont {Dan},
  \citenamefont {Geva},\ and\ \citenamefont {Batista}}]{intro_svd_2}%
  \BibitemOpen
  \bibfield  {author} {\bibinfo {author} {\bibfnamefont {X.}~\bibnamefont
  {Dan}}, \bibinfo {author} {\bibfnamefont {E.}~\bibnamefont {Geva}},\ and\
  \bibinfo {author} {\bibfnamefont {V.~S.}\ \bibnamefont {Batista}},\ }\href
  {https://arxiv.org/abs/2411.12049} {\bibinfo {title} {Qheom: A quantum
  algorithm for simulating non-markovian quantum dynamics using the
  hierarchical equations of motion}} (\bibinfo {year} {2024})\BibitemShut
  {NoStop}%
\bibitem [{\citenamefont {Schlimgen}\ \emph {et~al.}(2022)\citenamefont
  {Schlimgen}, \citenamefont {Head-Marsden}, \citenamefont {Sager},
  \citenamefont {Narang},\ and\ \citenamefont {Mazziotti}}]{intro_lcu_1}%
  \BibitemOpen
  \bibfield  {author} {\bibinfo {author} {\bibfnamefont {A.~W.}\ \bibnamefont
  {Schlimgen}}, \bibinfo {author} {\bibfnamefont {K.}~\bibnamefont
  {Head-Marsden}}, \bibinfo {author} {\bibfnamefont {L.~M.}\ \bibnamefont
  {Sager}}, \bibinfo {author} {\bibfnamefont {P.}~\bibnamefont {Narang}},\ and\
  \bibinfo {author} {\bibfnamefont {D.~A.}\ \bibnamefont {Mazziotti}},\
  }\bibfield  {title} {\bibinfo {title} {Quantum simulation of the lindblad
  equation using a unitary decomposition of operators},\ }\bibfield  {journal}
  {\bibinfo  {journal} {Physical Review Research}\ }\textbf {\bibinfo {volume}
  {4}},\ \href {https://doi.org/10.1103/physrevresearch.4.023216}
  {10.1103/physrevresearch.4.023216} (\bibinfo {year} {2022})\BibitemShut
  {NoStop}%
\bibitem [{\citenamefont {Schlimgen}\ \emph {et~al.}(2021)\citenamefont
  {Schlimgen}, \citenamefont {Head-Marsden}, \citenamefont {Sager},
  \citenamefont {Narang},\ and\ \citenamefont {Mazziotti}}]{original_paper}%
  \BibitemOpen
  \bibfield  {author} {\bibinfo {author} {\bibfnamefont {A.~W.}\ \bibnamefont
  {Schlimgen}}, \bibinfo {author} {\bibfnamefont {K.}~\bibnamefont
  {Head-Marsden}}, \bibinfo {author} {\bibfnamefont {L.~M.}\ \bibnamefont
  {Sager}}, \bibinfo {author} {\bibfnamefont {P.}~\bibnamefont {Narang}},\ and\
  \bibinfo {author} {\bibfnamefont {D.~A.}\ \bibnamefont {Mazziotti}},\
  }\bibfield  {title} {\bibinfo {title} {Quantum simulation of open quantum
  systems using a unitary decomposition of operators},\ }\href
  {https://doi.org/10.1103/PhysRevLett.127.270503} {\bibfield  {journal}
  {\bibinfo  {journal} {Phys. Rev. Lett.}\ }\textbf {\bibinfo {volume} {127}},\
  \bibinfo {pages} {270503} (\bibinfo {year} {2021})}\BibitemShut {NoStop}%
\bibitem [{\citenamefont {Suri}\ \emph {et~al.}(2023)\citenamefont {Suri},
  \citenamefont {Barreto}, \citenamefont {Hadfield}, \citenamefont {Wiebe},
  \citenamefont {Wudarski},\ and\ \citenamefont
  {Marshall}}]{two_unitary_decomposition}%
  \BibitemOpen
  \bibfield  {author} {\bibinfo {author} {\bibfnamefont {N.}~\bibnamefont
  {Suri}}, \bibinfo {author} {\bibfnamefont {J.}~\bibnamefont {Barreto}},
  \bibinfo {author} {\bibfnamefont {S.}~\bibnamefont {Hadfield}}, \bibinfo
  {author} {\bibfnamefont {N.}~\bibnamefont {Wiebe}}, \bibinfo {author}
  {\bibfnamefont {F.}~\bibnamefont {Wudarski}},\ and\ \bibinfo {author}
  {\bibfnamefont {J.}~\bibnamefont {Marshall}},\ }\bibfield  {title} {\bibinfo
  {title} {Two-unitary decomposition algorithm and open quantum system
  simulation},\ }\href {https://doi.org/10.22331/q-2023-05-15-1002} {\bibfield
  {journal} {\bibinfo  {journal} {Quantum}\ }\textbf {\bibinfo {volume} {7}},\
  \bibinfo {pages} {1002} (\bibinfo {year} {2023})}\BibitemShut {NoStop}%
\bibitem [{\citenamefont {Crooks}(2019)}]{theory_crooks_psr}%
  \BibitemOpen
  \bibfield  {author} {\bibinfo {author} {\bibfnamefont {G.~E.}\ \bibnamefont
  {Crooks}},\ }\bibfield  {title} {\bibinfo {title} {Gradients of parameterized
  quantum gates using the parameter-shift rule and gate decomposition}\ }\href
  {https://doi.org/10.48550/ARXIV.1905.13311} {10.48550/ARXIV.1905.13311}
  (\bibinfo {year} {2019})\BibitemShut {NoStop}%
\bibitem [{\citenamefont {Wierichs}\ \emph {et~al.}(2022)\citenamefont
  {Wierichs}, \citenamefont {Izaac}, \citenamefont {Wang},\ and\ \citenamefont
  {Lin}}]{general_psr}%
  \BibitemOpen
  \bibfield  {author} {\bibinfo {author} {\bibfnamefont {D.}~\bibnamefont
  {Wierichs}}, \bibinfo {author} {\bibfnamefont {J.}~\bibnamefont {Izaac}},
  \bibinfo {author} {\bibfnamefont {C.}~\bibnamefont {Wang}},\ and\ \bibinfo
  {author} {\bibfnamefont {C.~Y.-Y.}\ \bibnamefont {Lin}},\ }\bibfield  {title}
  {\bibinfo {title} {General parameter-shift rules for quantum gradients},\
  }\href {https://doi.org/10.22331/q-2022-03-30-677} {\bibfield  {journal}
  {\bibinfo  {journal} {Quantum}\ }\textbf {\bibinfo {volume} {6}},\ \bibinfo
  {pages} {677} (\bibinfo {year} {2022})}\BibitemShut {NoStop}%
\bibitem [{\citenamefont {Peetz}\ \emph {et~al.}(2024)\citenamefont {Peetz},
  \citenamefont {Smart},\ and\ \citenamefont {Narang}}]{peetz_scu}%
  \BibitemOpen
  \bibfield  {author} {\bibinfo {author} {\bibfnamefont {J.}~\bibnamefont
  {Peetz}}, \bibinfo {author} {\bibfnamefont {S.~E.}\ \bibnamefont {Smart}},\
  and\ \bibinfo {author} {\bibfnamefont {P.}~\bibnamefont {Narang}},\ }\href
  {https://doi.org/10.48550/arXiv.2407.21095} {\bibinfo {title} {Quantum
  {Simulation} via {Stochastic} {Combination} of {Unitaries}}} (\bibinfo {year}
  {2024}),\ \bibinfo {note} {arXiv:2407.21095 [quant-ph]}\BibitemShut {NoStop}%
\bibitem [{\citenamefont
  {Chakraborty}()}]{chakrabortyImplementingAnyLinear2023}%
  \BibitemOpen
  \bibfield  {author} {\bibinfo {author} {\bibfnamefont {S.}~\bibnamefont
  {Chakraborty}},\ }\href {http://arxiv.org/abs/2302.13555} {\bibinfo {title}
  {Implementing any {{Linear Combination}} of {{Unitaries}} on
  {{Intermediate-term Quantum Computers}}}},\ \Eprint
  {https://arxiv.org/abs/2302.13555} {2302.13555} \BibitemShut {NoStop}%
\bibitem [{\citenamefont {Nielsen}\ and\ \citenamefont
  {Chuang}(2010)}]{Nielsen_Chuang_2010}%
  \BibitemOpen
  \bibfield  {author} {\bibinfo {author} {\bibfnamefont {M.~A.}\ \bibnamefont
  {Nielsen}}\ and\ \bibinfo {author} {\bibfnamefont {I.~L.}\ \bibnamefont
  {Chuang}},\ }\href@noop {} {\emph {\bibinfo {title} {Quantum Computation and
  Quantum Information: 10th Anniversary Edition}}}\ (\bibinfo  {publisher}
  {Cambridge University Press},\ \bibinfo {year} {2010})\BibitemShut {NoStop}%
\bibitem [{\citenamefont {Childs}\ and\ \citenamefont
  {Wiebe}(2012)}]{LCU_Weibe}%
  \BibitemOpen
  \bibfield  {author} {\bibinfo {author} {\bibfnamefont {A.~M.}\ \bibnamefont
  {Childs}}\ and\ \bibinfo {author} {\bibfnamefont {N.}~\bibnamefont {Wiebe}},\
  }\bibfield  {title} {\bibinfo {title} {Hamiltonian simulation using linear
  combinations of unitary operations},\ }\href@noop {} {\bibfield  {journal}
  {\bibinfo  {journal} {Quantum Info. Comput.}\ }\textbf {\bibinfo {volume}
  {12}},\ \bibinfo {pages} {901–924} (\bibinfo {year} {2012})}\BibitemShut
  {NoStop}%
\bibitem [{\citenamefont {Hai}\ and\ \citenamefont
  {Ho}(2024)}]{lagrange_interpolation_psr}%
  \BibitemOpen
  \bibfield  {author} {\bibinfo {author} {\bibfnamefont {V.~T.}\ \bibnamefont
  {Hai}}\ and\ \bibinfo {author} {\bibfnamefont {L.~B.}\ \bibnamefont {Ho}},\
  }\bibinfo {title} {Lagrange interpolation approach for general
  parameter-shift rule},\ in\ \href
  {https://doi.org/10.1007/978-3-031-37966-6_1} {\emph {\bibinfo {booktitle}
  {Quantum Computing: Circuits, Systems, Automation and Applications}}},\
  \bibinfo {editor} {edited by\ \bibinfo {editor} {\bibfnamefont
  {H.}~\bibnamefont {Thapliyal}}\ and\ \bibinfo {editor} {\bibfnamefont
  {T.}~\bibnamefont {Humble}}}\ (\bibinfo  {publisher} {Springer International
  Publishing},\ \bibinfo {address} {Cham},\ \bibinfo {year} {2024})\ p.\
  \bibinfo {pages} {1–17}\BibitemShut {NoStop}%
\bibitem [{\citenamefont {Moler}\ and\ \citenamefont
  {Van~Loan}(1978)}]{19_dubious_ways_lagrange}%
  \BibitemOpen
  \bibfield  {author} {\bibinfo {author} {\bibfnamefont {C.}~\bibnamefont
  {Moler}}\ and\ \bibinfo {author} {\bibfnamefont {C.}~\bibnamefont
  {Van~Loan}},\ }\bibfield  {title} {\bibinfo {title} {Nineteen dubious ways to
  compute the exponential of a matrix},\ }\href
  {https://doi.org/10.1137/1020098} {\bibfield  {journal} {\bibinfo  {journal}
  {SIAM Review}\ }\textbf {\bibinfo {volume} {20}},\ \bibinfo {pages}
  {801–836} (\bibinfo {year} {1978})}\BibitemShut {NoStop}%
\bibitem [{\citenamefont {Tarantola}(2006)}]{Tarantola_2006}%
  \BibitemOpen
  \bibfield  {author} {\bibinfo {author} {\bibfnamefont {A.}~\bibnamefont
  {Tarantola}},\ }\href {https://doi.org/10.1007/978-3-540-31107-2} {\emph
  {\bibinfo {title} {Elements for Physics}}}\ (\bibinfo  {publisher} {Springer
  Berlin Heidelberg},\ \bibinfo {address} {Berlin, Heidelberg},\ \bibinfo
  {year} {2006})\BibitemShut {NoStop}%
\bibitem [{\citenamefont {Hailu}(2020)}]{lagrange_degen1}%
  \BibitemOpen
  \bibfield  {author} {\bibinfo {author} {\bibfnamefont {D.~H.}\ \bibnamefont
  {Hailu}},\ }\href {https://arxiv.org/abs/2004.05159} {\bibinfo {title} {On
  sylvester solution for degenerate eigenvalues}} (\bibinfo {year} {2020}),\
  \Eprint {https://arxiv.org/abs/2004.05159} {arXiv:2004.05159 [quant-ph]}
  \BibitemShut {NoStop}%
\bibitem [{\citenamefont {Richardson}\ and\ \citenamefont
  {Gaunt}(1927)}]{Richardson_Gaunt_1927}%
  \BibitemOpen
  \bibfield  {author} {\bibinfo {author} {\bibfnamefont {L.~F.}\ \bibnamefont
  {Richardson}}\ and\ \bibinfo {author} {\bibfnamefont {J.~A.}\ \bibnamefont
  {Gaunt}},\ }\bibfield  {title} {\bibinfo {title} {Viii. the deferred approach
  to the limit},\ }\href {https://doi.org/10.1098/rsta.1927.0008} {\bibfield
  {journal} {\bibinfo  {journal} {Philosophical Transactions of the Royal
  Society of London. Series A, Containing Papers of a Mathematical or Physical
  Character}\ }\textbf {\bibinfo {volume} {226}},\ \bibinfo {pages} {299–361}
  (\bibinfo {year} {1927})}\BibitemShut {NoStop}%
\bibitem [{\citenamefont {Krebsbach}\ \emph {et~al.}(2022)\citenamefont
  {Krebsbach}, \citenamefont {Trauzettel},\ and\ \citenamefont
  {Calzona}}]{richardson_error}%
  \BibitemOpen
  \bibfield  {author} {\bibinfo {author} {\bibfnamefont {M.}~\bibnamefont
  {Krebsbach}}, \bibinfo {author} {\bibfnamefont {B.}~\bibnamefont
  {Trauzettel}},\ and\ \bibinfo {author} {\bibfnamefont {A.}~\bibnamefont
  {Calzona}},\ }\bibfield  {title} {\bibinfo {title} {Optimization of
  richardson extrapolation for quantum error mitigation},\ }\href
  {https://doi.org/10.1103/PhysRevA.106.062436} {\bibfield  {journal} {\bibinfo
   {journal} {Phys. Rev. A}\ }\textbf {\bibinfo {volume} {106}},\ \bibinfo
  {pages} {062436} (\bibinfo {year} {2022})}\BibitemShut {NoStop}%
\bibitem [{\citenamefont {Fujiwara}(2004)}]{generalized_adc}%
  \BibitemOpen
  \bibfield  {author} {\bibinfo {author} {\bibfnamefont {A.}~\bibnamefont
  {Fujiwara}},\ }\bibfield  {title} {\bibinfo {title} {Estimation of a
  generalized amplitude-damping channel},\ }\href
  {https://doi.org/10.1103/PhysRevA.70.012317} {\bibfield  {journal} {\bibinfo
  {journal} {Phys. Rev. A}\ }\textbf {\bibinfo {volume} {70}},\ \bibinfo
  {pages} {012317} (\bibinfo {year} {2004})}\BibitemShut {NoStop}%
\bibitem [{\citenamefont {Rost}\ \emph {et~al.}(2020)\citenamefont {Rost},
  \citenamefont {Jones}, \citenamefont {Vyushkova}, \citenamefont {Ali},
  \citenamefont {Cullip}, \citenamefont {Vyushkov},\ and\ \citenamefont
  {Nabrzyski}}]{adc_krausmap}%
  \BibitemOpen
  \bibfield  {author} {\bibinfo {author} {\bibfnamefont {B.}~\bibnamefont
  {Rost}}, \bibinfo {author} {\bibfnamefont {B.}~\bibnamefont {Jones}},
  \bibinfo {author} {\bibfnamefont {M.}~\bibnamefont {Vyushkova}}, \bibinfo
  {author} {\bibfnamefont {A.}~\bibnamefont {Ali}}, \bibinfo {author}
  {\bibfnamefont {C.}~\bibnamefont {Cullip}}, \bibinfo {author} {\bibfnamefont
  {A.}~\bibnamefont {Vyushkov}},\ and\ \bibinfo {author} {\bibfnamefont
  {J.}~\bibnamefont {Nabrzyski}},\ }\bibfield  {title} {\bibinfo {title}
  {Simulation of thermal relaxation in spin chemistry systems on a quantum
  computer using inherent qubit decoherence},\ }\href
  {https://api.semanticscholar.org/CorpusID:226281597} {\bibfield  {journal}
  {\bibinfo  {journal} {arXiv: Quantum Physics}\ } (\bibinfo {year}
  {2020})}\BibitemShut {NoStop}%
\bibitem [{\citenamefont {Hu}\ \emph {et~al.}()\citenamefont {Hu},
  \citenamefont {Xia},\ and\ \citenamefont
  {Kais}}]{huQuantumAlgorithmEvolving2020a}%
  \BibitemOpen
  \bibfield  {author} {\bibinfo {author} {\bibfnamefont {Z.}~\bibnamefont
  {Hu}}, \bibinfo {author} {\bibfnamefont {R.}~\bibnamefont {Xia}},\ and\
  \bibinfo {author} {\bibfnamefont {S.}~\bibnamefont {Kais}},\ }\bibfield
  {title} {\bibinfo {title} {A quantum algorithm for evolving open quantum
  dynamics on quantum computing devices},\ }\href
  {https://doi.org/10.1038/s41598-020-60321-x} {\bibfield  {journal} {\bibinfo
  {journal} {Scientific Reports}\ }\textbf {\bibinfo {volume} {10}},\ \bibinfo
  {pages} {3301}}\BibitemShut {NoStop}%
\bibitem [{\citenamefont {Arrasmith}\ \emph {et~al.}(2020)\citenamefont
  {Arrasmith}, \citenamefont {Cincio}, \citenamefont {Somma},\ and\
  \citenamefont {Coles}}]{Arrasmith_Cincio_Somma_Coles_2020}%
  \BibitemOpen
  \bibfield  {author} {\bibinfo {author} {\bibfnamefont {A.}~\bibnamefont
  {Arrasmith}}, \bibinfo {author} {\bibfnamefont {L.}~\bibnamefont {Cincio}},
  \bibinfo {author} {\bibfnamefont {R.~D.}\ \bibnamefont {Somma}},\ and\
  \bibinfo {author} {\bibfnamefont {P.~J.}\ \bibnamefont {Coles}},\ }\bibfield
  {title} {\bibinfo {title} {Operator sampling for shot-frugal optimization in
  variational algorithms},\ }\href {http://arxiv.org/abs/2004.06252} {\
  (\bibinfo {year} {2020})},\ \bibinfo {note} {arXiv:2004.06252
  [quant-ph]}\BibitemShut {NoStop}%
\end{thebibliography}
\end{document}